\documentclass[a4paper,11pt]{article}
\pdfoutput=1 

\usepackage{jcappub} 

\usepackage[T1]{fontenc} 

\usepackage[utf8]{inputenc}
\usepackage{url}
\usepackage{hyperref}
\usepackage{tabularx}
\usepackage{amsmath}
\usepackage{amsfonts}
\usepackage{amssymb}
\usepackage{graphicx}
\graphicspath{ {./images/} }
\usepackage{float}
\usepackage[dvipsnames]{xcolor}
\usepackage{wrapfig}
\usepackage{subcaption}
\usepackage{wasysym}
\usepackage{appendix}
\usepackage[font=small]{caption}
\usepackage{soul}
\usepackage{booktabs}
\usepackage{siunitx}
\title{\boldmath  Weak lensing higher-order statistics to disentangle modified gravity and massive neutrinos}

\author[a,b,c,d]{A. Vadal\`{a}}
\author[a,b]{V.F. Cardone}
\author[e]{S. Vinciguerra}
\author[f,g]{F. Bouch\`{e}}
\author[h,i,j]{M. Baldi}
\author[i,j]{C. Giocoli}

\affiliation[a]{INAF - Osservatorio Astronomico di Roma, via Frascati 33, 00078 Monte Porzio Catone (Roma), Italy}
\affiliation[b]{INFN - Sezione di Roma, Piazzale Aldo Moro, 2 - c/o Dipartimento di Fisica, Edificio G.Marconi, I-00185 Rome, Italy}
\affiliation[c]{Tor Vergata University, Via della Ricerca Scientifica, 1 - c/o Dipartimento di Fisica, I-00133 Rome, Italy}
\affiliation[d]{Sapienza Universit\`{a} di Roma, Piazzale Aldo Moro, 2 - c/o Dipartimento di Fisica, Edificio E. Fermi, I-00185 Rome, Italy}
\affiliation[e]{Aix-Marseille Universit\'{e}, CNRS, CNES, LAM, Marseille, France}
\affiliation[f]{Scuola Superiore Meridionale, Via Mezzocannone 4, 80138,
Napoli, Italy}
\affiliation[g]{INFN - Sezione di Napoli, Via Cinthia 6, 80126, Napoli, Italy}
\affiliation[h]{Dipartimento di Fisica e Astronomia, Università di Bologna, Via Gobetti 93/2, 40129 Bologna, Italy}
\affiliation[i]{INAF - Osservatorio di Astrofisica e Scienza dello Spazio di
Bologna, Via Piero Gobetti 93/3, 40129 Bologna, Italy}
\affiliation[j]{INFN - Sezione di Bologna, Viale Berti Pichat 6/2, 40127 Bologna, Italy}

\emailAdd{alessandro.vadala@inaf.it}
\emailAdd{vincenzo.cardone@inaf.it}
\emailAdd{simone.vinciguerra@lam.fr}
\emailAdd{filippo.bouche-ssm@unina.it}
\emailAdd{marco.baldi5@unibo.it}
\emailAdd{carlo.giocoli@inaf.it}

\abstract{
Going beyond second order in weak lensing (WL) statistics is known to break degeneracies among cosmological parameters. We take a step further here, investigating whether higher-order statistics (HOS) in weak lensing can disentangle among General Relativity (GR) and modified gravity (MG), also taking into account the presence of massive neutrinos. To this end, we rely on mock convergence maps obtained from GR and $f(R)$ gravity N\,-\,body simulations, and we look for MG signatures in a wide set of higher-order WL probes. We rely on different metrics to quantify the discriminatory power of each probe, also varying the measurement setup. We find out that WL HOS can indeed disentangle MG and GR also in the presence of massive neutrinos.
}

\begin{document}
\maketitle
\flushbottom

\section{Introduction}
\label{sec:intro}

The current standard cosmological model, referred to as $\Lambda$CDM, inherits its name from the two main components it assumes the Universe is dominated by. The cosmological constant $\Lambda$ fuels the observed cosmic speed up \cite{riess1998, perlmutter1999}, while the cold dark matter (CDM) plays a fundamental role in determining the hierarchical growth of structures and their clustering properties. Its success in fitting a wide set of observations probing both the kinematics and the dynamics of the Universe has motivated its adoption as concordance model \cite{planck2018, eBOSS_2021}. Notwithstanding the fascinating property of explaining the data with a minimal set of only six parameters, the $\Lambda$CDM model is nevertheless not free of theoretical shortcomings related to the lack of direct detection of any CDM particle, and of a successful interpretation of the $\Lambda$ origin (see, e.g., \cite{joyce2015} for a review).

These problems have been considered minor issues as far as the success of $\Lambda$CDM model in matching the data still stands out as the strongest motivation in favor of it. However, this point has started to be questioned over the years by the rise of cosmological tensions (see, e.g., \cite{divalentino2021}). The most striking one is, for sure, the Hubble tension, i.e., the difference between the value of the present-day Hubble constant $H_0$ as estimated from cosmological probes, such as the cosmic microwave background (CMB) data from Planck \cite{planck2018}, or those from the Baryon Oscillation Spectroscopic Survey (BOSS) \cite{boss2013}, and the one measured from the distance ladder methods such as local SNeIa \cite{riess2019}. The two $H_0$ values turn out to disagree at the $\sim 5 \sigma$ level, with the offset hard to explain in terms of unaccounted systematics in the data. Less severe but yet troublesome is the second, the tension about the clustering parameter $S_8 = \sigma_8 \sqrt{\Omega_\text{m} / 0.3}$ (see, e.g., \cite{perivalaropoulos2022}). Weak lensing and galaxy clustering probes (e.g, \cite{kids2013,des2020}) are in $\sim 3 \sigma$ tension with the Planck results \cite{planck2018}, although some recent results \cite{kids2025} point at systematics correction as a way to greatly alleviate the discrepancy.

The cracks within $\Lambda$CDM have sparked the investigation of possible alternatives and solutions. In particular, it is worth wondering whether such cracks can be taken as evidence for the need to question the implicit assumptions inside the model. The validity of General Relativity (GR) as the correct theory of gravitation on cosmological scales is the first cornerstone one could drop entering the vast realm of modified gravity (MG) theories \cite{Capozziello:2011et, clifton2012_mgreview}. The study of MG models has received a renewed interest as Stage\,IV surveys, like the Dark Energy Spectroscopic Instrument (DESI) \cite{desi2016}, the Legacy Survey of Space and Time (LSST) \cite{lsst2009}, \textit{Euclid} \cite{euclid2024overview}, and the Nancy Roman Space Telescope \cite{spergel2015}, have started or will start soon collecting data.

Weak lensing (hereafter, WL) has soon emerged as an ideal tool to discriminate among GR and MG models. It is, indeed, a probe of both the background expansion and the growth of structures so that it can break the degeneracy due to the possibility of MG models to perfectly match the kinematical quantities, e.g., the Hubble rate, while still predicting different lensing observables. The cosmological information from the WL signal is extracted primarily with two-point statistics, like the shear two-point correlation function, which are sufficient when studying Gaussian fields. However, the clustering of matter due to gravitational collapse introduces non-Gaussian features in the matter distribution, which might be missed by standard statistics \cite{BERNARDEAU20021_spt}. This motivates the use of higher-order statistics (HOS), which allow to have access to the wealth of information embedded in the non-Gaussianity of the lensing convergence field. The \textit{Euclid} Higher-Order Weak Lensing Statistics (HOWLS) team \cite{euclid_howls2023} has conducted an exploratory investigation to highlight how the constraints on the cosmological parameters can be narrowed down through the joint use of 2nd and higher-order WL probes. The analysis has also confirmed how HOS, although in general being more noisy, have the unique capability of breaking degeneracies between cosmological parameters, like $\Omega_{\text{m}}$ and $\sigma_8$ involved in the aforementioned $S_8$ tension.

The HOWLS analysis has been carried out in the framework of GR, but the application of HOS is not limited to dark energy models. On the contrary, peak counts, which are a subset among the HOS, have already been used to discriminate among rival MG theories \cite{Peel2018, davies2024}. In addition, other studies based on the same set of simulations used here, and employing convolutional neural networks, have demonstrated that incorporating statistics beyond the two‑point level into the feature vector increases the ability to discriminate between gravity models \cite{merten2019, peel2019}. The aim of the present paper is to extend the above results, considering a wider range of HOS probes, all of them already explored in the HOWLS paper. We will indeed rely on higher-order moments (HOMs) \cite{gatti_des2022}, peak statistic, measured on convergence and aperture mass maps ($M_{\text{ap}}$) \cite{martinet2021b}, one-point probability distribution function (1\,-\,PDF) \cite{barthelemy2020}, Betti numbers (BNs) \cite{Feldbrugge_2019}, and Minkowski functionals (MFs) \cite{Vicinanza2019, Parroni2020}. 

As a particular case of MG model, we consider $f(R)$ gravity, in which the dependence on the Ricci scalar $R$ in the Einstein-Hilbert action is extended from linearity to a generic function $f(R)$. Although it seems quite a trivial modification, it has been extensively studied due to its interesting phenomenology (see, e.g., \cite{sotiriou_fr2010}). We will also include the effect of massive neutrinos since it has been shown that their effect can compensate for the one of $f(R)$, thus making the departure from GR less evident \cite{baldi_deg2014}. As a way to make this test quantitative, we will measure the same probes on both GR and MG datasets obtained by measuring them on mock convergence maps built starting from the set of DUSTGRAIN-\textit{pathfinder} N-body simulations suite \cite{giocoli_dustgrain2018}. These have been carried out both in $\Lambda$CDM and in MG with different configurations, for a total of eight combinations of MG and neutrino mass parameters, and one reference  $\Lambda$CDM cosmology. \\

The plan of the paper is as follows. We will first present, in Sec.\,\ref{sec:modgrav}, the basics of $f(R)$ gravity, highlighting the expected signatures of departures from GR. The simulations suite is briefly introduced in Sec.\,\ref{sec:data}, while Sec.\,\ref{sec:stats} presents the HOS probes we will rely on. The metrics used to quantify the difference among different models, and the way we use them, are discussed in Sec.\,\ref{sec:discriminating}. Results are presented in Secs.\,\ref{sec:metrics_results} and \ref{sec:nonparametric_results}, while we summarize and conclude in Sec.\,\ref{sec:conclusion}. Additional material is presented in Appendices for completeness.

\section{$f(R)$ gravity}
\label{sec:modgrav}

From a mathematical point of view, MG theories differ from GR because of a different action postulated to assign the gravity Lagrangian. In the case of our interest, this is achieved by promoting the dependency on Ricci scalar in the Einstein-Hilbert action to a generic function of $R$. The class of theories thus obtained is popularly known as $f(R)$ {\it gravity}, also referred to as {\it fourth-order gravity} since it leads to fourth-order differential equations. Following the notation in \cite{STAROBINSKY1980}, the modified action reads (in units with $c = 1$)

\begin{equation}
\label{eq:fr_action}
    S = \int \text{d}^{4}x \sqrt{-g} \left[ \frac{R + f(R)}{16 \pi G_{N}} + \mathcal{L}_{m} \right]
\end{equation}
where $R$ is the Ricci (or curvature) scalar, $f(R)$ a generic function of $R$ (differentiable at least up to the second order), $G_{N}$ the Newtonian gravitational constant, and $\mathcal{L}_m$ is the Lagrangian of the matter field. Varying the action in Eq. \ref{eq:fr_action} with respect to the metric, we get the modified Einstein equations

\begin{equation}
\label{eq:fr_einsteineqs}
    G_{\mu \nu} + f_{R} R_{\mu \nu} - \left( \frac{f(R)}{2} - \Box f_{R}  \right) g_{\mu \nu} - \nabla_{\mu} \nabla_{\nu} f_{R} = 8 \pi G_{N} T_{\mu \nu}
\end{equation}
with $R_{\mu \nu}$ the Ricci tensor, $G_{\mu \nu} = R_{\mu \nu} - \frac{1}{2}g_{\mu \nu}R$ the Einstein one, $\Box = \nabla_{\alpha} \nabla^{\alpha}$ the covariant Laplacian, and $T_{\mu \nu}$ the stress-energy tensor, and a label $R$ denotes the derivative with respect to $R$, i.e., $f_{R} = {\rm d}f(R)/{\rm d}R$. Inserting the Robertson\,-\,Walker metric into Eqs.\, \ref{eq:fr_einsteineqs}, we get the modified Friedmann equations, which can be rearranged in a convenient way, showing that the effect of $f(R)$ is the same as that of a fluid with negative pressure. It is, indeed, this feature that makes it possible for $f(R)$ theories to match the data on the expansion of the Universe. Moreover, the higher degree of the equations allows to perfectly mimic the same background expansion, e.g., the same Hubble rate $H(z)$, as a given dark energy model, while still producing a modified growth of structures \cite{Capozziello:2005ku}.

Alternatively, the MG effects of $f(R)$ gravity can also be considered equivalent to the ones of an extra scalar field. Taking the trace of Eqs.\,\ref{eq:fr_einsteineqs}, we get

\begin{equation}
    \Box f_{R} = \frac{1}{3} [R - f_{R} R + 2 f(R) + 8 \pi G_{N} \rho] = \frac{\partial V_{\text{eff}}}{\partial f_{R}}
\end{equation}
that is indeed a Klein-Gordon field equation for the additional scalar degree of freedom $f_{R}$. Note that we have here collected all the terms on the right-hand side as the derivative of an effective potential $V_{eff}(R)$. The effective scalar field mediates an additional fifth force of gravitational origin, whose range is given by the Compton wavelength of the field itself $\lambda_{f_R} = m_{f_R}^{-1}$, having defined $m^2_{f_R} = \partial^2 V_{\text{eff}}/\partial f_R^2$.

Although there is a large freedom in the choice of the functional form of $f(R)$, there are nevertheless some constraints which must be fulfilled in order to both recover GR on the Solar System scale, and avoid instability in the field equations \cite{Bertotti2003, defelice2017}. Among the possible models passing this preliminary selection, we focus here on the model introduced by Hu \& Sawicki (HS) \cite{hu_sawicki2007} setting

\begin{equation}
\label{eq:hu_saw_fr}
    f(R) = -m^2 \frac{c_{1} \left( \frac{R}{m^2}\right)^n}{c_{2} \left( \frac{R}{m^{2}} \right)^n +1}
\end{equation}
where $m^{2}= \Omega_{m} H^2$ is the mass scale, $\Omega_{m}$ is the present value of the matter density parameter, $H$ the Hubble rate, while $(c_{1}, c_{2}, n)$ are model parameters that can be adjusted to fix the expansion history (with $n > 0$ in general). Assuming the background expansion is the same as for a flat $\Lambda$CDM, we can express the curvature as

\begin{equation}
    R \simeq 3 m^{2} \left( a^{-3} + 4 \frac{\Omega_{\Lambda}}{\Omega_{m}}\right)
\end{equation}
so that, setting $n = 1$ as for the case of our interest later, the scalar field takes the value

\begin{equation}
    f_{R} \simeq - \frac{c_{1}}{c_{2}^2} \left( \frac{m^2}{R} \right)^{2}
\end{equation}
showing that the amplitude of deviations from GR is set by $c_{1}/c_{2}^2$. It is then convenient to parameterize the HS model through the value of the scalar field today, which is given by

\begin{equation}
    f_{R0} = -\frac{1}{c_{2}} \frac{6\Omega_{\Lambda}}{\Omega_{m}} \left( \frac{m^2}{R_{0}} \right)^2.
\end{equation}
In general, it is straightforward to check that $\lim_{R \rightarrow 0}{f(R)} = 0$. This regime is achieved at large $z$ so that the deviations from GR turn off, thus not spoiling down the agreement with the strong constraints from the cosmic microwave background \cite{planck2018}. In the intermediate redshift range, when structure formation takes place, the wavelength of the perturbations becomes smaller than the Compton wavelength, thus turning on the deviations from GR. This causes an enhancement of the gravitational potential and an increase in the growth rate of structures, bringing deviations in the power spectrum with respect to GR, potentially detectable by weak lensing experiments. Finally, at smaller scales, the effect of $f(R)$ is reduced due to the screening effect known as \textit{chameleon mechanism} \cite{Capozziello:2007eu, Khoury_chameleon2013}. In this case, since the mass of the scalar field depends both on the potential and the coupling to the matter fields, it becomes environment dependent. The increase in the background density also increases the mass of the scalar degree of freedom, strongly reducing the range of the fifth force and bringing $f(R)$ back to standard gravity, thus passing the tight constraints on GR at small scales \cite{Bertotti2003, will2014}.

It is instructive to look at how $f(R)$ affects the lensing observables from a theoretical point of view. We start from the modified Poisson equations for Newtonian potential $\Psi$ and for the Weyl (lensing) potential $\Phi_+ = (\Phi + \Psi)/2$, in Fourier space

\begin{align}
\label{eq:mu_sigma}
    k^2 \Psi &= -4 \pi G_N a^2 \, \mu(a,k) \, \bar{\rho} \Delta \\
    k^2(\Phi+\Psi) &= -8 \pi G_N a^2 \, \Sigma(a,k) \, \bar{\rho} \Delta.
\end{align}
where the functions $\Sigma(a,k)$, $\mu(a,k)$, and $\eta(a,k) = \Phi / \Psi$ \cite{planck_2015_de_mg} can be assigned phenomenologically or computed from the given model. In the case of $f(R)$ gravity, it is

\begin{equation}
    \label{eq:sigma_mg}
    \Sigma(a) = \frac{1}{1+f_R(a)}
\end{equation}
so that one gets $\Sigma(a) \simeq 1$ with great accuracy since, for all viable $f(R)$ models, one has $f_R(a) << 1$. We can therefore conclude that the propagation of light in $f(R)$ is the same as in GR, but it is the clustering properties to differ, hence altering the WL observables due to the changes in the lensing convergence field. Motivated by this argument, we therefore look at statistics measured on convergence maps since it is here that $f(R)$ theories imprint their signature. 

\section{Simulations}
\label{sec:data}

While waiting for Stage\,IV data, a possible way to investigate whether WL HOS can discriminate among GR and $f(R)$ would be to compare the predictions from the two different scenarios. As pointed out, e.g., in the HOWLS paper \cite{euclid_howls2023}, a theoretical estimate is available only for some of the HOS probes, and often in idealized conditions which are rarely met in actual observations. To circumvent this problem, one can directly compare the HOS quantities as measured from mock data obtained from simulations carried out under different GR and MG models. This is the approach we will take in this paper.

\begin{table}[]
\centering
$
\begin{array}{lcccccc}
\hline \text { Simulation name } & \text { Gravity type } & f_{R 0} & m_v(\mathrm{eV}) & \Omega_{\mathrm{CDM}} & \Omega_v & \sigma_8 \\
\hline \Lambda \text{CDM } & \mathrm{GR} & - & 0 & 0.31345 & 0 & 0.847 \\
\rm{f R 4} & f(R) & -1 \times 10^{-4} & 0 & 0.31345 & 0  & 0.967 \\
\rm{f R 5} & f(R) & -1 \times 10^{-5} & 0 & 0.31345 & 0  & 0.903 \\
\rm{f R 6} & f(R) & -1 \times 10^{-6} & 0 & 0.31345 & 0  & 0.861  \\
\rm{f R 4} \_0.3 \mathrm{eV} & f(R) & -1 \times 10^{-4} & 0.3 & 0.30630 & 0.00715 & 0.893 \\
\rm{f R 5} \_0.15 \mathrm{eV} & f(R) & -1 \times 10^{-5} & 0.15 & 0.30987 &0.00358 & 0.864 \\
\rm{f R 5} \_0.1 \mathrm{eV} & f(R) & -1 \times 10^{-5} & 0.1 & 0.31107 & 0.00238 & 0.878 \\
\rm{f R 6} \_0.1 \mathrm{eV} & f(R) & -1 \times 10^{-6} & 0.1 & 0.31107 & 0.00238 & 0.836 \\
\rm{f R 6} \_0.06 \mathrm{eV} & f(R) & -1 \times 10^{-6} & 0.06 & 0.31202 & 0.00143 & 0.847 \\
\hline
\end{array}
$
\caption{Parameters of the different cosmologies in DUSTGRAIN-\textit{pathfinder} simulations}
\label{tab:fr_cosmo}
\end{table}

A preliminary consideration has guided the choice of the simulation suite to use. We are interested in finding the signatures of $f(R)$ gravity on the growth of structures as traced by WL observables. However, it is well known that, in many MG models, MG effects can be counterbalanced by those of massive neutrinos \cite{Motohashi2013, baldi_deg2014, Wright_2017}. In the case of $f(R)$ gravity, the boost in clustering due to the action of the effective fifth force can be compensated by the free streaming of massive neutrinos, which, on the contrary, suppresses that same clustering. It is then possible to find suitable combinations of the $f_{R0}$ parameter and the total neutrino mass, $M_{\nu} = \sum_i m_{\nu,i}$, leading to an almost perfect cancellation of the two opposite effects. As a consequence, two-point statistics probes, such as the convergence power spectrum or the halo mass function, may not be able to distinguish between $\Lambda$CDM and $f(R)$ when the neutrino mass is not known from independent observations \cite{baldi_deg2014,Peel2018}.

This consideration highlights the need for simulations that take into account the effect of massive neutrinos, and vary both $f_{R0}$ and $M_{\nu}$. This is why we have relied on the DUSTGRAIN\footnote{Dark Universe Simulations to Test GRAvity In the presence of Neutrinos}\,-\,{\it pathfinder} simulations. These are run for both GR and $f(R)$ gravity, adopting the HS model with $n = 1$ and $(c_1, c_2)$ set in such a way that the same background expansion as $\Lambda$CDM is obtained. In particular, this is set to be in agreement with \cite{Planck2015_cosmo} hence fixing: $\Omega_m = \Omega_{\text{CDM}} + \Omega_b + \Omega_{\nu} = 0.31345$, $\Omega_b = 0.0481$, $\Omega_{\Lambda} = 0.68655$, $H_0 = 67.31$ km s$^{-1}$ Mpc$^{-1}$, $\mathcal{A}_s = 2.199 \times 10^{-9}$, $n_s = 0.9658$ thus giving $\sigma_{8,0} = 0.847$. Massive neutrinos are included, and different choices are available for $(f_{R0}, M_{\nu}$). The simulations span the range $-1 \times 10^{-4} \leq f_{R0} \leq -1 \times 10^{-6}$ and 0 eV $\leq M_{\nu} \leq$ 0.3 eV. We here considered the cosmologies with the parameter combinations as they appear in \cite{giocoli_dustgrain2018}, that is 8 $f(R)$ simulations with different values of the neutrino mass plus an additional $\Lambda$CDM simulation with standard GR taken as reference. Table\,\ref{tab:fr_cosmo} summarizes the parameter values for the different cosmologies adopted.

The DUSTGRAIN-\textit{pathfinder} simulations include 768$^{3}$ dark matter particles with $m_{CDM} = 8.1 \times 10^{10} M_{\astrosun} h^{-1}$ for $M_{\nu} = 0$ and an equivalent number of neutrinos for the case of $M_{\nu} \neq 0$, within a box of side 750 Mpc $h^{-1}$, following the dynamics dictated by $f(R)$ equations of motion. Past light cones are reconstructed from slicing 21 comoving snapshots with $0 < z < 4$ and, with a squared footprint of 5 deg per side, distributing the particles onto 27 different lens planes depending on the comoving distance from the observer and whether they lie in the field of view. In the routine adopted, the observer is placed at the vertex of a pyramid whose square base is at the comoving distance corresponding to $z=4$.

We had at our disposal 256 different light-cone realizations, obtained by randomizing the comoving cosmological boxes for each cosmology. Each convergence map has a resolution of $2048 \times 2048$ pixels, which means a pixel angular resolution of 8.8 arcsec.

The lensing planes are constructed by storing, for each plane $l$, on each pixel of coordinate $(i,j)$, the particle mass density

\begin{equation}
    \Xi_l(i,j)  = \frac{\sum_k m_k}{A_l}
\end{equation}
where $A_l$ is the comoving pixel area of the $l$-lens plane and $\sum_k m_k$ the total mass of the particles in the pixel.
Finally, the convergence map at a given source redshift is computed as

\begin{equation}
    \kappa = \sum_l \frac{\Xi_l}{\Xi_{\text{crit},l,s}}
\end{equation}
where the sum is extended over the $l$ lens planes with redshift smaller than the source redshift $z_s$, while $\Xi_{\text{crit},l,s} = (c^2/4 \pi G)(D_L/D_S D_{LS})$ is the critical surface density at the lens plane $l$ for sources at $z_s$, being $D_L$, $D_S$, $D_{LS}$ the observer-lens, observer-source, and source-lens angular diameters distances.

For each line of sight (LOS), we load the simulated convergence ($\kappa$) maps generated from the underlying lensing planes and first subtract their mean to enforce $\langle \kappa \rangle = 0$. To resemble the setup expected from upcoming surveys such as \textit{Euclid}, we downsample each field onto a $512 \times 512$ grid (corresponding to $\simeq 0'.59$ per pixel), averaging the signal in the new wider pixels.  The injection of the shape noise follows Eq. 15 of \cite{Castiblanco_2024}, where we generate and add a pure shape noise realization with intrinsic ellipticity dispersion per component $\sigma_\epsilon = 0.26$ and a galaxy number density $n_{\rm gal}=30\,\mathrm{arcmin}^{-2}$, using a random seed tied to the LOS index to ensure reproducibility and to mimic random variations of the noise properties from real observations.
These processed maps, corresponding to source redshifts $z_s \in [0.5,1.0,2.0,4.0]$, and generated assuming the same shape noise realization, are used as inputs for the subsequent statistical analysis. The choice of assigning a constant $n_{\mathrm{gal}}$ at all source redshift planes, although is not intended to be fully realistic, is adopted to isolate the signatures of MG with respect to GR without introducing confounding observational effects related to the source redshift distribution. This is consistent with the scope of this work, which does not aim to reproduce any specific real survey.

\section{Summary statistics}
\label{sec:stats}

We attempt to break the degeneracy between modified gravity and massive neutrino effects by using a variety of probes beyond the 2PCF, which can be categorized into two classes. On the one hand, we consider one-point statistics, including the 1\,-\,PDF and HOM, which provide information on the distribution of the projected matter density field and its low-order moments. On the other hand, we use a set of probes, namely $\kappa$\,- and $M_{\mathrm{ap}}$\,-\,peaks, MFs, and BNs, which can be collectively referred to as topological statistics, as they characterize the distribution of critical points in the field and how their connectivity varies with the signal-to-noise ratio. The need for such a broad suite of tools is motivated both by the sensitivity of these probes to different features of the field and by the comparable performance of the HOS in extracting the non-Gaussian information encoded in the WL field (see, e.g., \cite{euclid_howls2023}), which prevents us from favoring a single statistic or a limited subset of them.

The statistical tools we use in this work already have an extensive literature behind them (see, e.g., \cite{martinet2018,harnois_deraps2021,gatti_des2024}). In the following, we will therefore only sketch their definitions, and what they probe referring to \cite{euclid_howls2023} and refs. therein for further details. In particular, we will not address how to compute theoretical predictions (when possible) since we will rely on direct measurements.

\subsection{Two\,-\,point correlation function}

Although our main interest is in HOS, we also include the standard two\,-\,points statistics both for completeness, and as a comparison to show the better discriminatory power of HOS. As a 2nd order probe, we consider the convergence 2\,-\,points correlation function (2PCF)

\begin{equation}
\xi_{\kappa}(\theta) = \langle \kappa(\boldsymbol{\vartheta}) \kappa(\boldsymbol{\vartheta} + \boldsymbol{\theta}^{\prime}) \rangle
\label{eq: xikappadef}
\end{equation}
where, in the left-hand side, we have taken into account the homogeneity and isotropy conditions to replace the vector $\boldsymbol{\theta}$ with its magnitude $\theta$. Its counterpart in the harmonic space is the convergence power spectrum defined through

\begin{equation}
\label{eq:gamma_kappa}
    \langle \tilde{\kappa}(\boldsymbol{\ell})\tilde{\kappa}^*(\boldsymbol{\ell}') \rangle = (2 \pi)^2 \delta_D (\boldsymbol{\ell} - \boldsymbol{\ell}') C_{\kappa} (\ell)
\end{equation}
where $\delta_D$ is the Dirac delta function and the 2D wave vector $\boldsymbol{\ell}$ is the Fourier conjugate of $\boldsymbol{\theta}$, and we have again used the isotropy and homogeneity conditions to get $C_{\kappa}(\boldsymbol{\ell}) = C_{\kappa}(\ell)$.

It is useful to remind that, in the harmonic space, shear and convergence are related by

\begin{equation}
    \tilde{\gamma}(\boldsymbol{\ell}) = \frac{(\ell_2 + i \ell_2)^2}{\ell^2}\tilde{\kappa}(\boldsymbol{\ell}) = e^{2i\beta} \tilde{\kappa}(\boldsymbol{\ell})
\end{equation}
which means that the power spectra of convergence and shear coincide, $C_{\gamma} = C_{\kappa}$.

Since the convergence is a scalar quantity, we have only one 2PCF, and it is then easy to show that it is $\xi_{\kappa}(\theta) = \xi_{+}(\theta)$. We can therefore measure the convergence 2PCF either on the convergence map itself or on the shear one. The consistency between the two sets of measurements can be taken as evidence that there are no unaccounted systematics affecting the map-making process. Since we have direct access to the mock convergence maps, in this paper we will rely on direct measurements on them to estimate $\xi_{\kappa}(\theta)$.

The 2PCF $\xi_{\kappa}$ is measured using \texttt{TreeCorr}\footnote{\url{https://github.com/rmjarvis/TreeCorr}} \cite{jarvis_2004}. We employ logarithmic angular bins with a minimum separation set by the pixel scale, $\theta_{\min}=0'.59$, and a maximum separation corresponding to the map diagonal, $\simeq 427'$, using a total of 50 bins.

\subsection{One\,-\,point probability distribution function}

The late time density field can be efficiently characterized through the one\,-\,point convergence probability distribution function (1\,-\,PDF).
On the smaller scales dominated by shape noise, a preliminary smoothing is needed in order to make the variance of the clustering probed by the 1\,-\,PDF larger than that of the shape noise itself. In this limit, this quantity can be straightforwardly measured, or its information can be compressed in related summary statistics such as the density split \cite{Friedrich2018des}. Another attractive feature of 1\,-\,PDF is that, on mildly nonlinear scales, an exact theoretical formulation is possible \cite{barthelemy2020,boyle2021_1pdf}. However, it is worth stressing that, in order for the theoretical derivation to work, one has to use smoothing radii $\sim 10 \ {\rm arcmin}$ which, at low redshift, indeed translate (in the Limber approximation) to scales $k$ in the mildly nonlinear regime. However, in this paper, we are not interested in a theoretical modeling so that we can push the smoothing radius to smaller values.

To measure the 1\,-\,PDF of the convergence field, we start from the processed noisy $\kappa$ maps, each smoothed using a top-hat\footnote{\url{https://docs.astropy.org/en/stable/api/astropy.convolution.Tophat2DKernel.html}} filter of 4 increasing angular scales, $(2.4, 4.7, 9.4, 18.9)$ arcmin,  applied via FFT convolution. The current filtering is meant to ease the level of shape noise at the cost of partially losing non-Gaussian information on the small scales. To avoid edge effects, we exclude pixels within a distance equal to the smoothing radius from the map boundaries. The mean of the smoothed map is subtracted to ensure $\langle \kappa \rangle = 0$.  The 1\,-\,PDF is then computed by binning the pixel values into a uniform grid of 201 midpoints spanning $\kappa \in [-0.1, 0.1]$, with corresponding bin edges constructed to fully explore the under- and over-dense tails of the distribution.  The resulting histogram, normalized to unity, provides the PDF of $\kappa$ for each smoothing scale.

\subsection{Higher\,-\,order moments}

As the 2PCF and harmonic space power spectrum are related to the variance of the field, one can push the correlation statistics to more points. Limiting to no more than four points, one can therefore look at the 3\,- and 4\,-\,points CF, or to their harmonic space counterparts, namely bispectrum and trispectrum, respectively. As an alternative, one can rely on the associated compressed statistics, which will then characterize the skewness and kurtosis of the convergence field. This is our choice here, taking as observables the higher\,-\,order moments (HOM) up to the fourth order, denoting them as $\langle \kappa^n \rangle$. with $n = (2, 3, 4)$.

As for the 1\,-\,PDF, we first need to smooth the convergence field, and then just compute the mean value of $\kappa_{s}^{n}(\theta_s)$ with $\kappa_s(\theta_s)$ the smoothed convergence as a function of the smoothing radius $\theta_s$. One could indeed use $\langle \kappa^n \rangle = \langle \kappa^n \rangle(\theta_s)$ as quantities to be matched by the theory. However, we found out that the measurements at different $\theta_s$ values are strongly correlated, so that the constraining power can be satisfactory shown by taking a limited number $\theta_s$ values.

To compute the moments of the convergence field, we follow the analogous smoothing strategy as described above for the 1\,-\,PDF using a top-hat kernel.  After subtracting the mean, we compute the
central moments of the smoothed field, namely $\langle \kappa^2 \rangle$ (variance), $\langle \kappa^3 \rangle$ (skewness), and $\langle \kappa^4 \rangle$ (kurtosis), by averaging the respective powers of the pixel values.

Note that the smoothing procedure does not completely remove the noise so that the HOM are still contaminated, and this must be taken into account in the modeling. One could perform a further denoising by subtracting suitable combinations of noise moments estimated from noise only maps so that the agreement with the theoretical predictions improves \cite{gatti_des2022, Vicinanza2019, Parroni2020}. Since we are not interested here in any modeling, we will skip this step.

\subsection{Peak statistics}

An over\,-\,density along the line of sight causes a peak in the convergence field, i.e., it manifests as a region where the signal\,-\,to\,-\,noise ratio (S/N) is larger than in the surroundings. It is intuitive to understand that large S/N peaks are signatures of the presence of galaxy clusters. On the other hand, random superpositions of many minor overdensities can also originate peaks in the convergence map with intermediate S/N values. In this case, it is the line-of-sight structure itself to act as a single overdensity. It should be noted, however, that detection of low-amplitude peaks might be due also to shape noise contamination, therefore resulting unsensitive to a change in cosmology \cite{martinet2018}. From these arguments, it is therefore clear that cosmological information from non-Gaussian features can also be extracted from the peaks of convergence maps. As the name suggests, studying WL peaks means studying the distribution of local maxima in over-dense regions as a function of the S/N. We will consider the full S/N range\footnote{A detailed modeling is available only in the large S/N regime, where clusters are the main cause for the presence of peaks (see, e.g., \cite{kruse2000, bartelmann2001}). As we are not interested in modeling, this is not a problem for our aims.} since preliminary studies based on Fisher matrix technique have shown that this choice maximizes the constraining power \cite{euclid_howls2023}. A filter function is needed in order to extract peaks from the convergence maps due to the presence of the galaxy shape noise. Through the convolution with a filter function on a given scale, it is possible to measure the peak in the maps at a specific scale. An efficient way of doing it is by first transforming the convergence map into an aperture mass map, $M_{\mathrm{ap}}$ \cite{Schneider1998_apm}. One of the main advantages in using $M_{\mathrm{ap}}$ over a filtered $\kappa$ field is the convolution with a compensated filter function, which better focuses on frequencies around the selected scale, making more optimal the peaks detection. For a given sky location $\boldsymbol{\theta}$, we define 

\begin{equation}
\label{eq:apm}
    M_{\text{ap}}(\boldsymbol{\theta};\vartheta) = \int d^2 \theta' \, U_{\vartheta}(|\boldsymbol{\theta} - \boldsymbol{\theta'}|) \, \kappa(\boldsymbol{\theta'})
\end{equation}
where $U_{\vartheta}(\boldsymbol{\theta})$ is a compensated filter function defined over a circle angular radius $\boldsymbol{\vartheta}$, vanishing for $\boldsymbol{\theta} > \boldsymbol{\vartheta}$, or at least going to zero smoothly for small $\boldsymbol{\theta}$. $M_{\text{ap}}$ can also be expressed in terms of the shear, provided the filter $U_{\vartheta}$ is compensated, i.e., it must satisfy the condition

\begin{equation}
\label{eq:comp_filter}
    \int d^2 \theta \, U_{\vartheta}(\boldsymbol{\theta}) = 0.
\end{equation}
Aperture mass peak counts are measured from convergence maps smoothed with a compensated filter, namely the JBJ04 \cite{jarvis_2004}, at 3 angular scales $(2'.4,  4'.7,  9'.4)$ using Fourier-space convolution\footnote{\url{https://github.com/sheydenreich/threepoint/blob/main/python_scripts/aperture_mass_computer.py}}. We drop the 4th smoothing radius at $18'.9$ as the number of features is too low, thus making measurements extremely noisy. Pixels within a distance equal to the smoothing radius from the map boundaries are excluded to avoid edge effects.  After subtracting the mean and normalizing the smoothed map by its standard deviation, local maxima (i.e., peaks) are detected as pixels higher than their 8 neighbors. The number of peaks is then binned according to their signal-to-noise ratio $\nu = M_{\mathrm{ap}}/\sigma_{M_{\mathrm{ap}}}$ using 150 bins in the range $M_{\mathrm{ap}} \in [-5, 10]$, and counts at the map borders are
excluded. This procedure is repeated for each LOS and smoothing scale.

\subsection{Betti numbers}

In a generic field, non-Gaussianities are imprinted not only on its clustering properties, but also on its topology. Topology is the branch of mathematics that deals with the properties of objects preserved under continuous transformations. It addresses the identity of topological features and their connectivity; that is, how field regions with a value exceeding a given threshold, encompassing a certain number of cavities, are interconnected through tunnels.

A relevant role in topology is played by the Betti numbers \cite{Betti1870,Feldbrugge_2019}, which count the features in terms of the number of $p$-dimensional holes. We refer the reader to \cite{edelsbrunner2010, Bray2021, Vick1994} for a formal explanation, while a shorter introduction can also be found in \cite{Feldbrugge_2019}.

Essentially, BNs indicate the number of all linearly independent topological cycles. In particular, it is possible to show that a $N$-dimensional manifold is characterized by $N+1$ BNs, so that for the convergence field, we have three BNs ($\beta_0$, $\beta_1$, $\beta_2$). In particular, $\beta_0$ counts the number of disjoint components, while $\beta_1$ gives the number of non\,-\,equivalent loops enclosing troughs. Finally, it is possible to show that $\beta_2$ is identically null. 

We can then see what happens when varying the threshold value; decreasing $\nu$, the introduction of a vertex increases $\beta_0$ by 1, the introduction of a face decreases $\beta_1$ by 1, while the introduction of an edge either connects two parts, decreasing $\beta_0$, or introduces a cycle, increasing $\beta_1$. $\beta_2$ increases only when the lowest minimum is added. This is the basis of the algorithm to measure $(\beta_0, \beta_1)$ from the convergence maps as function of the threshold $S/N$.

The first two Betti numbers, $\beta_{0}$ and $\beta_{1}$, are measured from the top-hat smoothed convergence maps using the code validated in \cite{parroni_2021}. Each patch is recentered to zero mean and normalized by its standard deviation, yielding the signal-to-noise field $\nu = \kappa/\sigma_\kappa$.  Betti numbers are detected over a uniform grid of thresholds $\nu \in [-5,5]$ with spacing $\Delta\nu = 0.05$, for a total of 201 midpoints.  For each threshold, the excursion set $\{\nu > \nu_i\}$ is identified, and $\beta_{0}(\nu_{i})$ is computed as the number of connected components.  The complementary set $\{\nu \le \nu_i\}$ is used to identify holes, from which $\beta_{1}(\nu_{i})$ is obtained by counting
connected regions that do not intersect the boundary.  The resulting Betti curves $\beta_{0}(\nu)$ and $\beta_{1}(\nu)$ provide a topological summary of the weak-lensing field\footnote{Betti numbers are computed using \texttt{scikit-image} routines, including \texttt{label} and \texttt{regionprops}; \url{https://scikit-image.org/docs/stable/api/skimage.measure.html\#skimage.measure.label},
\url{https://scikit-image.org/docs/stable/api/skimage.measure.html\#skimage.measure.regionprops}.}.

\subsection{Minkowski functionals}

As a further topological probe, we consider the Minkowski functionals  \cite{adler2009} since they have already proven to be a reliable tool in cosmology \cite{Schmalzing1995_mink,Schmalzing1998}. Moreover, their application to convergence maps has already been explored \cite{Vicinanza2019,Parroni2020,Munshi2021_mfs}.

It might be more intuitive to rely on the geometrical interpretation of the MFs. To this end, let us consider a smooth 2D random field $f(\mathbf{x})$ with zero mean and variance $\sigma_0^2$. We first introduce the excursion set $Q_{\nu} = \{ \mathbf{x} | f(\mathbf{x})/ \sigma_0 \geq \nu \}$, that is the region where the normalized field $f/\sigma_0$ is larger than a given threshold $\nu$. We can then define the integral form of the first three MFs as

\begin{subequations} \label{mfs_equations}
\begin{align}
    V_0(\nu) &= \frac{1}{A} \int_{Q_{\nu}} da \\
    V_1(\nu) &= \frac{1}{4A} \int_{\partial Q_{\nu}} dl \\
    V_2(\nu) &= \frac{1}{2\pi A} \int_{\partial Q_{\nu}} dl \, \mathcal{K}
\end{align}
\end{subequations}
where $A$ is the area of the map on which the measurement is performed, $\partial Q_{\nu}$ is the boundary of the excursion set (where the expression in $Q_{\nu}$ is satisfied by equality), $da$ and $dl$ are, respectively, the surface and line elements along $\partial Q_{\nu}$, while $\mathcal{K}$ is the local geodesic curvature of $\partial Q_{\nu}$. According to these definitions, we can interpret $V_0$ and $V_1$ as, respectively, the area and the perimeter of $Q_{\nu}$, while $V_2$ is the Euler characteristic of $Q_{\nu}$ (i.e., the difference between the number of disconnected regions above and below the threshold $\nu$). 

Similarly to the BNs, the three MFs of the convergence field are measured from the top-hat smoothed and normalized maps, using the signal-to-noise field $\nu = \kappa / \sigma_\kappa$ with the same binning strategy as the former probe. Making use of the code proposed in \cite{Vicinanza2019}, we evaluate the area, perimeter, and integrated curvature of the excursion sets $\{\nu > \nu_i\}$, by computing the numerical discrete integrals in Eqs.\,\ref{mfs_equations}.  This yields the MF triplet $\{V_0(\nu), V_1(\nu), V_2(\nu)\}$, which characterizes the morphology of the WL convergence field at each smoothing scale.

\section{Discriminating between GR and MG}
\label{sec:discriminating}

The effective scalar field of $f(R)$ theories changes the clustering of matter on the nonlinear scales probed by the different HOS we have introduced above. A comparison among the same summary statistics measured on simulations with the same background but different growth should therefore provide evidence for deviations from GR. The presence of noise can, however, masks out such evidence so that it is mandatory to perform the comparison on convergence maps with realistic noise added. This is particularly true for HOS, which are very sensitive to shape noise. Moreover, even if a difference is spotted out, one needs methods to quantify whether it is statistically meaningful or not. We will rely on two different approaches, which we describe in the following paragraphs. Before moving further, we note that we assumed the source galaxies to lie on fixed redshift planes, with $z_s \in [0.5, 1.0, 2.0, 4.0]$. Since no redshift distribution has been adopted, our results are not representative of any specific survey, but instead provide clear insights into the potential of HOS in modified gravity scenarios and into what can be expected from Stage IV surveys.

\subsection{Gaussianity test}

Although in the following we will not perform a likelihood analysis, it is worth noticing that previous investigations about the use of HOS assumed that the likelihood itself can be modeled as Gaussian. This same assumption will be useful in the following when we will quantify the discriminatory power for some of the methods. As already shown in the literature \cite{euclid_howls2025}, the HOS data vector must be suitably cut in order for the Gaussianity assumption be fulfilled. To this end, we use the procedure described below.

\begin{enumerate}
\item[i.]{We first rebin the data vectors to enhance the signal-to-noise ratio. Consecutive bins are grouped together by taking their mean (for 1\,-\,PDF and MFs) or their sum (for peaks, aperture mass peaks, and Betti numbers). No rebinning is applied to the 2PCF. The results are independent of the bin width as long as extreme under- or over-sampling of the data vectors is avoided.} \\

\item[ii.]{We then compute
    \begin{equation}
        y_i = \left(\textbf{D}_{i} - \langle \textbf{D} \rangle \right)^\mathrm{T} \textbf{C}^{-1} \left(\textbf{D}_{i} - \langle \textbf{D} \rangle \right) \, ,
    \end{equation}
where $\textbf{D}_{i}$ is the data vector from the i-th simulation, $\langle \textbf{D} \rangle$ is the mean over all realizations, and \textbf{C} is the numerical covariance matrix estimated from a dedicated set of DUSTGRAIN-\textit{pathfinder} GR simulations. Under the Gaussian likelihood assumption, the vector \textbf{y} is expected to follow a $\chi^2$ distribution with degrees of freedom equal to the length of $\textbf{D}_{i}$. However, when the length of $\textbf{D}_{i}$ approaches the number of simulations used to estimate the numerical covariance, the $\chi^2$ is no longer an appropriate reference, and an empirical target distribution becomes necessary.} \\

\item[iii.]{From a multivariate Gaussian distribution with mean $\langle \textbf{D} \rangle$ and covariance $\textbf{C}$, we sample 500 Gaussian realizations of the dataset, each consisting of 256 data vectors $\textbf{D}_{i}$. For each realization, we compute the probability distribution $P_{\mathrm{gauss}}\left(y_i\right)$. Then, we take the mean of $P_{\mathrm{gauss}}\left(y_i\right)$ across the 500 Gaussian realizations as our empirical target distribution.} \\

\item[iv.]{We compute the weighted Symmetrized Mean Absolute Percentage Error (SMAPE, \cite{Rizzato:2022hbu}) between \textbf{y} and the target distribution,
    \begin{equation}\label{eq:smape}
        \mathcal{S}_{\mathrm{obs}} = \frac{1}{\sum_i \omega_i} \sum_i \omega_i \frac{\left| P_{\mathrm{obs}}\left(y_i\right) - \langle P_{\mathrm{gauss}} \left(y_i\right) \rangle \right|}{\left|    P_{\mathrm{obs}}\left(y_i\right) \right| + \left| \langle P_{\mathrm{gauss}} \left(y_i\right) \rangle \right|} \, ,
    \end{equation}
where $P_{\mathrm{obs}}\left(y_i\right)$ is the measured probability distribution of $y_i$, $\langle P_{\mathrm{gauss}} \left(y_i\right) \rangle$ is the reference empirical distribution, and the weights $\omega_i = \langle P_{\mathrm{gauss}} \left(y_i\right) \rangle$ ensure that the central region of the distribution dominates the SMAPE estimate, reducing the influence of deviations in the tails.} \\

\item[v.]{We compare the observed SMAPE value to the empirical threshold $\mathcal{S}_{\mathrm{lim}}$, which can be computed from the 500 Gaussian realizations that have been drawn in step (iii). Specifically, for each Gaussian realization, we compute the SMAPE value according to Eq.~\ref{eq:smape}. Then, we derive the mean and standard deviation across the 500 estimates and define the threshold as
\begin{equation}
    \mathcal{S}_{\mathrm{lim}} = \langle \mathcal{S}_{\mathrm{obs}}^{\mathrm{gauss}} \rangle + 2 \sigma \left( \mathcal{S}_{\mathrm{obs}}^{\mathrm{gauss}} \right) \, .
\end{equation}} \\

\item[vi.]{If ${S}_{\mathrm{obs}} < {S}_{\mathrm{lim}}$, the Gaussian likelihood assumption is fulfilled. Otherwise, we cut data vectors by removing the non-Gaussian tail elements and repeat steps (ii)-(v) until the Gaussian regime is achieved.}
\end{enumerate}
It is worth noticing that there are different choices that can make a given probe pass the Gaussianity test. On the other hand, should one be unable to find one, we will discard the probe for reasons that will be explained later. This does not mean that the probe can not be used, but only that can not be fit in our framework.

\subsection{Metrics}
\label{sec:metrics}

As a first approach, we look at the distance between the data vectors measured on GR and $f(R)$ simulated convergence maps. The issue here is to explain what we consider as {\it distance}. Four options are considered as briefly explained below, referring the reader to, e.g., \cite{feigelson2012} for further details. In the following, we denote with $({\bf D}_{GR}, {\bf D}_{MG})$ the set of measurements of a given HOS on the $\Lambda$CDM, and on a given $f(R)$\,+\,massive neutrino model defined by assigning the values of $(f_{R0}, M_{\nu})$. How the data vector is built depends on the HOS probe of interest and will be discussed in the results section.

The easiest metric we consider is the SMAPE itself  \cite{euclid_howls2023,flores1986_smape} already used in the Gaussianity test. We redefine it as 

\begin{equation}
{\cal{S}}_{NW} = \frac{1}{{\cal{N}}_{data}} \sum_{i = 1}^{{\cal{N}}_{data}} \frac{100 \left | D_{GR}^{(i)} - D_{MG}^{(i)} \right |}{\left ( \left | D_{GR}^{(i)} \right | + \left | D_{MG}^{(i)} \right | \right )/2} \ ,
\label{eq: smapenwdef}
\end{equation}
where the sum is over the ${\cal{N}}_{data}$ elements $D^{(i)}$ of the data vector. The closer is ${\cal{S}}$ to zero, the more the two datasets are indistinguishable. One can, however, argue that the points with the larger S/N should weigh more. Since the error is estimated from the same data covariance matrix, the points with the larger S/N are those with the larger signal. One can therefore define a weighted version of the SMAPE as

\begin{equation}
{\cal{S}}_{YW} = \left \{ \sum_{i = 1}^{{\cal{N}}_{data}} \frac{100 \left | D_{GR}^{(i)} \right |^2 \left | D_{GR}^{(i)} - D_{MG}^{(i)} \right |}{\left ( \left | D_{GR}^{(i)} \right | + \left | D_{MG}^{(i)} \right | \right )/2} \right \} 
\left \{ \sum_{i = 1}^{{\cal{N}}_{data}} \left | D_{GR}^{(i)} \right |^2 \right \}^{-1} \ .
\label{eq: smapeywdef}
\end{equation}
Note that the choice of the GR datavector as weight is arbitrary, but has no impact on our results since we are interested in comparing SMAPE values rather than the individual ones.

Since we have 256 realizations of each data vector, we compute the non-weighted and weighted SMAPE, $({\cal{S}}_{NW}, {\cal{S}}_{YW})$, for each one, and look at the mean values denoted as $(\langle {\cal{S}}_{NW} \rangle, \langle {\cal{S}}_{YW} \rangle)$ to quantify the discriminatory power between GR and MG of the given HOS.

The comparison of the average SMAPE values does not take into account the errors in the data. These are, on the contrary, explicitly taken care of in the definition of the other metrics we will use. The first one is the Mahalanobis distance \cite{mahalanobis1936generalized} given by

\begin{equation}
d_M = \sqrt{({\bf D}_{GR} - {\bf D}_{MG})^{{\rm T}}
{\bf C}^{-1} ({\bf D}_{GR} - {\bf D}_{MG})}
\label{eq: mahadef}
\end{equation}
which quantifies the distance between the GR and MG data vectors weighted by the data covariance matrix ${\bf C}$. Note that here ${\bf D}$ denotes the final estimate of the data vector after averaging over the 256 realizations, while ${\bf C}$ is estimated from a dedicated sample, and assumed to be cosmology independent. It is worth noticing that, under the assumption of Gaussianity, the Mahalanobis distance enters the likelihood being ${\cal{L}} \propto (1/\sqrt{\det(2\pi \mathbf{C}})\exp(-d_M^2/2)$.

Actually, one could have estimated the covariance matrix for both the GR and MG datasets separately and account for any discrepancy between the two. This can be done using the Wasserstein distance\cite{kantorovich1960_weisserstein} defined as

\begin{equation}
d_{W}^{2} = \Vert {\bf D}_{GR} - {\bf D}_{MG} \Vert^2 + \text{Tr}\left[ \mathbf{C}_{GR} + \mathbf{C}_{MG} - 2 \left( \mathbf{C}_{GR}^{\frac{1}{2}} \mathbf{C}_{MG} \mathbf{C}_{GR}^{\frac{1}{2}} \right)^{\frac{1}{2}} \right]
\label{eq: wassdef}
\end{equation}
where $\Vert {\bf x} \Vert^2$ is the vector magnitude, and $\mathbf{C}_i$ are the respective covariance matrices of the two models. Although we could estimate the covariance of the GR and MG models from the simulations, this is not what is done in practice when working with survey data. On the contrary, one first estimate ${\bf C}_{GR}$, and then assumes to be model independent so that ${\bf C}_{MG} = {\bf C}_{GR}$. In this case, the second term on the right-hand side of Eq.(\ref{eq: wassdef}) vanishes, and the Wasserstein metric reduces to the Euclidean distance between the two vectors. This simplified definition is the one we will use in the present work.

As a final metric, we also consider the Hellinger distance \cite{Hellinger1909}, which is used to quantitatively describe the similarity between the probability distribution functions. This reads

\begin{equation}
d_{H}^{2} = 1 - 
\sqrt{\frac{2 | {\bf C}_{GR} |^{1/2} | {\bf C}_{MG} |^{1/2}}
{| {\bf C}_{GR} + {\bf C}_{MG} |}}
\exp{\left [-
\frac{({\bf D}_{GR} - {\bf D}_{MG})^{\rm T} 
({\bf C}_{GR} + {\bf C}_{MG})^{-1} 
({\bf D}_{GR} - {\bf D}_{MG})}{4} \right ]}
\label{eq: helldef}
\end{equation}
where $|{\bf C}|$ is the determinant of the matrix ${\bf C}$. We will again assume the covariance to be the same for the two samples so that $\sigma_{GR} = \sigma_{MG}$. Noticing that it is $\mu_{i} ={\bf D}_i$, the Hellinger distance, in this case, reduces to

\begin{equation}
d_H^{2} = 1 - \exp{(-d_M^2/8)}
\label{eq: dhdm}
\end{equation}
so that one could drop this metric. We prefer to retain it since the presence of the exponential makes it more sensitive to small differences than the Mahalanobis distance.

It is straightforward to check that all four metrics introduced here go to zero for ${\bf D}_{GR} = {\bf D}_{MG}$. One could then naively take a non-null value as evidence for MG. However, both data vectors have been estimated from a fairly small number of realizations, so that the estimate of each metric is noisy. As such, it is possible that two different sets of a finite number of GR realizations do not provide the same estimate for a given metric ${\cal{M}}$. Put in other words, we must attach an error to ${\cal{M}}$ so that we can consider a difference $\Delta {\cal{M}} = {\cal{M}}_{GR} - {\cal{M}_{MG}} \neq 0$ as evidence for deviations from GR only if

\begin{displaymath}
\left | \Delta {\cal{M}} \right | > 3 \sigma(\Delta {\cal{M}}) \simeq 3 \left [ \sigma^2({\cal{M}}_{GR}) + \sigma^2({\cal{M}}_{MG}) \right ]^{1/2}
\end{displaymath}
with $\sigma({\cal{M}})$ the error on ${\cal{M}}$, and we have assumed uncorrelated errors on the right-hand side. We then proceed as schematically sketched below.

\begin{enumerate}

\item[i.]{For a given metric ${\cal{M}} \in \{{\cal{S}}_{NW}, {\cal{S}}_{YW}, d_M, d_W, d_H\}$ and an observable probe ${\cal{O}}$ (2PCF, 1\,-\,PDF, peaks, aperture mass peaks, Betti numbers, MFs), we first compute the fiducial data vector ${\bf D}_{GR}({\cal{O}})$ and ${\bf D}_{MG}({\cal{O}})$ from the 256 DUSTGRAIN noisy convergence maps.} \\

\item[ii.]{We now generate 256 data vectors ${\bf D}_{GR}^{rnd}({\cal{O}})$ and ${\bf D}_{MG}^{rnd}({\cal{O}})$ sampling from Gaussian distribution centered on the fiducial values computed before and with covariance as estimated from the DUSTGRAIN GR simulations.} \\

\item[iii.]{For each realization, we compute the metric ${\cal{M}}$ replacing in the corresponding definition ${\bf D}_{MG}$ with either ${\bf D}_{GR}^{rnd}({\cal{O}})$ or ${\bf D}_{MG}^{rnd}({\cal{O}})$ to get ${\cal{M}}_{GR}^{rnd}$ and ${\cal{M}}_{MG}^{rnd}$, respectively.} \\

\item[iv.]{We repeat steps i.\,-\,iii. 500 times, and look at the set of $({\cal{M}}_{GR}^{rnd}, {\cal{M}}_{MG}^{rnd})$ values thus obtained for the GR and MG cases to finally estimate $({\cal{M}}_{GR}, {\cal{M}}_{MG})$ and their errors as the median and symmetrized $68\%$ confidence ranges of the corresponding distributions.}

\end{enumerate}
Note that, in step ii., we have sampled from Gaussian distributions, which is the same as assuming that the data do follow a Gaussian distribution. This is indeed the case since we have performed a cut on the full data vector for each HOS probe in order to make them pass the Gaussianity test previously described. Since HOM do not pass this test, we will not analyze them with the metrics approach presented in this paragraph.

\subsection{Nonparametric tests}
\label{subsec: nonparamstests}

For fixed model and choice of measurement setup, we have measured each HOS probe on the 256 DUSTGRAIN realizations so that we have not only a fiducial data vector, but a full distribution. We can therefore go beyond the comparison of the final data vectors by investigating whether the GR and MG distributions are in agreement or not. To this end, we rely on nonparametric hypothesis testing (see, e.g., \cite{feigelson2012}). These methods allow us to quantitatively tell if a statistical significant difference occurs between the measurement of the same HOS in different models, without the need to make assumptions about the underlying distribution of the given probe. The hypothesis tests we considered in this work are the following:

\begin{itemize}

\item{Kolmogorov\,-\,Smirnov (KS) \cite{smirnov1948} based on on the maximum distance between the empirical distribution function (e.d.f.) of two samples;} \\

\item{Cram\'{e}r\,-\,von Mises (CvM) \cite{cramer1928} relying on the sum of the squared differences between the e.d.f. of two samples;} \\

\item{Ansari-Bradley (AB) \cite{ansari1960} estimating the ratio of the variances of the samples;} \\

\item{Mann-Whitney U (MWU) \cite{mannwhitney1947} testing if the two samples are drawn from the same distribution, computing the sum of ranks.}
\end{itemize}

\begin{figure} [H]
    \centering
    \includegraphics[width=0.8\linewidth]{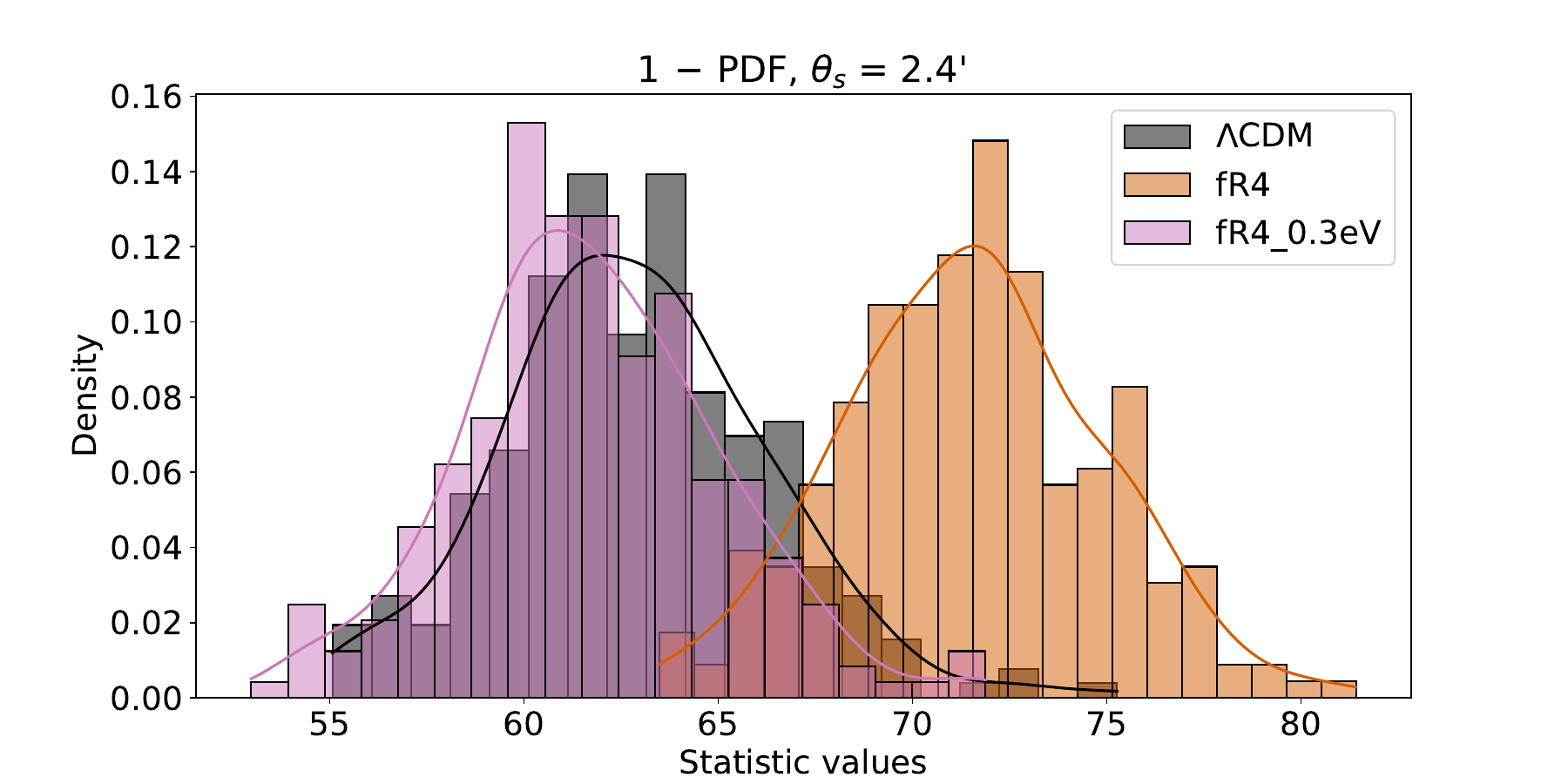}
    \caption{Histograms of samples of a selected bin, at $z_s=2.0$, smoothing $2'.4$, for the 1\,-\,PDF, obtained by counting the values of the statistic from the 256 maps of the DUSTGRAIN simulations falling into the given bin. The different distributions correspond respectively to $\Lambda$CDM, and the fR4 cosmologies with and without massive neutrinos. The solid lines come from the Gaussian smoothing obtained with KDE. The histograms are normalized so the total area equals unity.}
    \label{fig:example-histogram}
\end{figure}

In order to use the above tests, we first need to arrange the measurements in samples that we can compare. To this end, for a given choice of model and configuration setup (i.e., smoothing scale, range, redshift), we split the range $({\cal{O}}_{min}, {\cal{O}}_{max})$ in 20 bins\footnote{We have checked that results do not significantly change if we use more bins.}, and, for each bin, make the histogram of the 256 realizations of that bin. We thus get 20 different histograms which we apply a Gaussian smoothing to using a kernel density estimation (KDE), which provide an estimate of the underlying PDF of each sample. An example of the outcome is shown in Fig.\,\ref{fig:example-histogram} for the 1\,-\,PDF measured at $z_s = 2.0$ with a $2.4 \ {\rm arcmin}$ smoothing.

All the methods listed above are nonparametric hypothesis tests under the null hypothesis, $h_0$, that the two samples are drawn from the same underlying (and unknown) distribution. In addition to the raw value of the test statistic, we evaluate the corresponding p-value $p$, adopting a quite conservative significance level of $\alpha = 0.001$, to be highly confident not to observe false positives due to noise fluctuations.

When we evaluate the test statistic, if the p-value satisfies $p < \alpha$, we can reject $h_0$, and we observe a statistically significant difference between the two samples at the chosen confidence level; on the other hand, if $p > \alpha$, we fail to reject $h_0$ and we are unable to discriminate between the two samples obtained from the cosmological models considered.

The formal computation of the p-value depends on the specific test statistic of interest (e.g., for the CvM test, the method in \cite{anderson1962} is used). We relied on the \texttt{scipy.stats} \cite{virtanen2020_scipy} module in \texttt{Python} to perform the tests and compute the associated p-values. Those tests are repeated, in each configuration, for every bin, keeping in the results only those that passed the test, that is the ones satisfying the condition $p < \alpha$.

\section{Results from metrics-based method}
\label{sec:metrics_results}

There are different choices that must be made before computing the metrics introduced above. Since they are based on comparing data vectors, one must first choose, for each probe, how the data vector itself is assembled. Given the source redshift, and the cosmological model, and denoting with ${\cal{O}}(x)$ a generic HOS probe, one has to select a range for the $x$ variable (being the angular distance $\theta$ for the 2PCF, the convergence $\kappa$ for the 1\,-\,PDF, or the S/N threshold for peaks and topological probes), the number of $x$ bins, and the smoothing radius $\theta_s$. In the following, We will refer to each particular choice of range\,+\,binning\,+\,smoothing as configuration or setup. By varying these ingredients, we have explored a large number of configurations only retaining those passing the Gaussianity tests. This leaves us with still too large number of cases, so that, for each fixed $z_s$, we select the 25 ones giving the data vectors with the largest total S/N, estimated as ${\rm S/N} = \left ( {\bf D}^{\rm T} {\rm C}^{-1} {\bf D} \right )^{1/2}$ with ${\bf C}$ the data covariance matrix. We then use the procedure described in Sec.\,\ref{sec:metrics} to evaluate each metric ${\cal{M}}$ and the associated uncertainty in both GR and MG scenarios of interest for the 100 configurations thus selected. We finally denote with $f_{mod}({\cal{O}}, {\cal{M}})$ the fraction of cases with $\Delta {\cal{M}}/\sigma({\cal{M}}) > 3$ for the probe ${\cal{O}}$ and the model $mod$. We consider this quantity both without splitting in $z_s$, and for the four source redshift values. We thus have a quantitative and concise way to compare different probes for a given model, the larger $f_{mod}({\cal{O}}, {\cal{M}})$ being evidence for a stronger discriminatory power.

\begin{table}
\centering
$
\begin{array}{cccccccc}
\hline 
{\cal{M}} \ \  & \ \ {\rm 2PCF} & \ \ {\rm 2PCC} & \ \ {\rm 1\,-\,PDF} & \ \ {\rm \kappa\,-\,peaks} & \ \ {M_{\mathrm{ap}}\,-\,\rm{peaks}} & \ \ {\rm BNs} & \ \ {\rm MFs} \\
\hline
{\cal{S}}_{NW} & 0.0 & 0.0 & 55.0 & 0.0 & 20.2 & 0.0 & 11.8 \\
{\cal{S}}_{YW} & 0.0 & 0.0  & 56.0 & 2.1 & 54.8 & 0.0 & 9.7 \\
d_M & 89.0 & 0.0 & 55.0 & 14.4 & 75.0 & 21.3 & 47.3 \\
d_W & 0.0 & 0.0 & 56.0 & 1.0 & 69.0 & 0.0 & 9.7 \\
d_H & 89.0 & 0.0 & 55.0  & 14.4 & 75.0 & 21.3 & 47.3 \\
\hline
\end{array}
$
\caption{Values of $f_{mod}({\cal{O}}, {\cal{M}})$ for different probes ${\cal{O}}$ and metrics ${\cal{M}}$ taking the ${\rm fR5\_0.10eV}$ model as reference and summing over the four source redshifts.}
\label{tab: fmodvalref}
\end{table}

Note that this quantity depends also on the metric ${\cal{M}}$ adopted since the difference in the data vectors could be better highlighted by one metric rather than another on a case\,-\,by\,-\,case basis. As an example, we report in Tab.\,\ref{tab: fmodvalref} the values of $f_{mod}({\cal{O}}, {\cal{M}})$ choosing the ${\rm fR5\_0.10eV}$ model, and varying both the probe ${\cal{O}}$ and the metric ${\cal{M}}$ summing the results for all source redshifts. Ideally, the metric most suitable to discriminate among GR and MG is the one which, for a given probe, gives the largest $f_{mod}({\cal{O}}, {\cal{M}})$ values. Taken at face values, the numbers in the 4th to 8th columns point at the Mahalanobis and Hellinger distances as the preferred choice. We warn the reader that the fact that $f_{mod}$ is the same for both $d_M$ and $d_H$ metric is related to the fact that they are actually not independent on each other. However, if we look at the $f_{mod}$ values for fixed $z_s$, a difference can be seen with the Hellinger distance typically performing best. It is also worth noticing that, should we change the model and/or we look at the results at fixed $z_s$, a different ranking could sometimes be present depending on the HOS probe. However, the Hellinger distance performs best in most of the cases, or stays in the top three positions so that we will hereafter report results for this metric only. Should we adopt a different metric, the conclusions would nevertheless be qualitatively similar.

An important remark concerns columns 2 and 3 in Tab.\,\ref{tab: fmodvalref}. Here, 2PCF refers to the results obtained using the 2pt correlation function as measured on the convergence map. Taking the Hellinger distance as a reference metric, one surprisingly finds that the 2PCF works better than any HOS probe which is in striking contradiction with what we have said up to now. We find a similar result also for other $f(R$) models, and if we split the sources according to their redshift. This is actually an outcome of unrealistic assumptions. Indeed, while HOS probes are measured after smoothing the map with a filter of size $\theta_s$, the 2PCF is measured on the unsmoothed map. As a consequence, the 2PCF is able to trace the power spectrum also on scales smaller than $\theta_s$, which are, on the contrary, removed from the HOS probes because of the smoothing. Since these are the highly nonlinear scales where the difference between GR and MG is maximized, the discriminatory power of the 2PCF is artificially boosted. In practice, such small scales are removed from the analysis since they are heavily affected by the uncertainties in the modeling of baryons. In order to make a fair comparison with HOS, we have therefore considered what we refer to as 2PCC in the 3rd column. Here, we use the 2PCF but cutting scales smaller than $\theta_s$. The $f_{mod}$ values drop to zero, thus confirming that the discriminatory power was originating by the unrealistic assumption of using all scales in the 2PCF. We have checked that this holds true also for other cases so that we will not report anymore the results for either 2PCF or 2PCC.

\begin{figure}[htbp]
    \centering
    
    \begin{subfigure}[b]{0.48\textwidth}
        \centering
        \includegraphics[width=\linewidth]{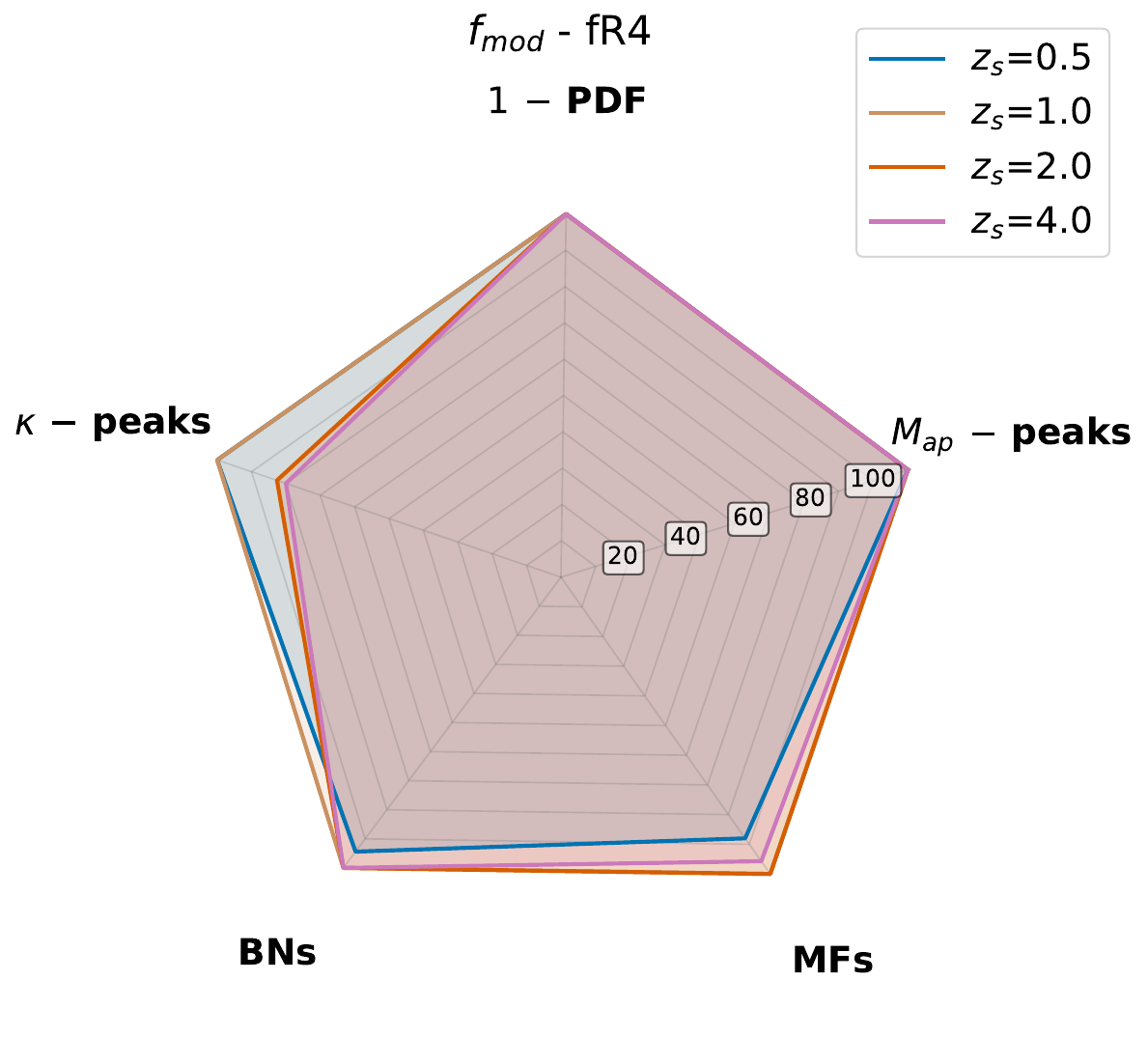}
    \end{subfigure}
    \begin{subfigure}[b]{0.48\textwidth}
        \centering
        \includegraphics[width=\linewidth]{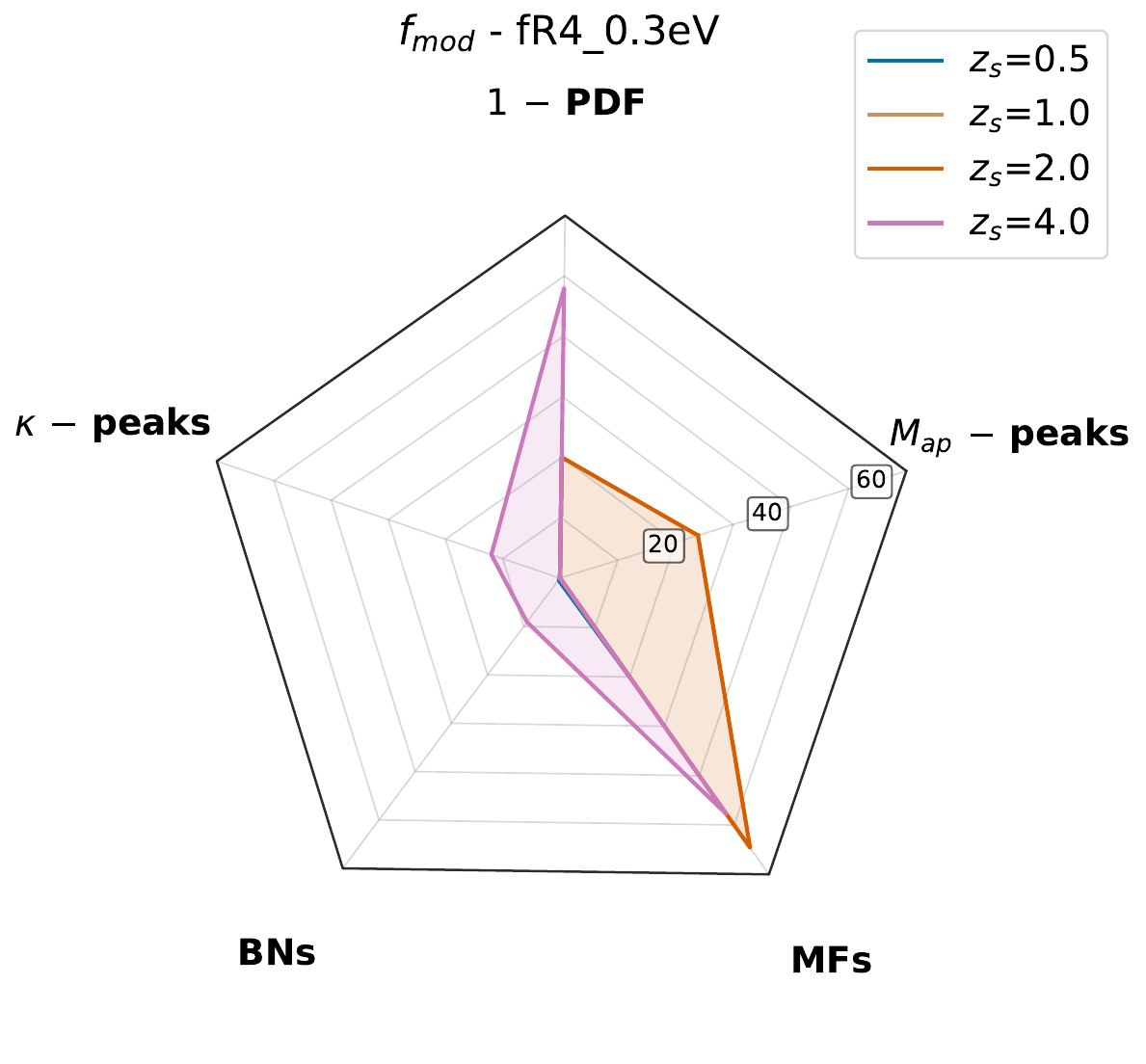}
    \end{subfigure}
    \caption{Values of $f_{mod}({\cal{O}}, {\cal{M}})$ as a function of the source redshift $z_s$ using the Hellinger distance as metric for fR4 cosmologies. Please note the scale change in the left panel for better viewing.}
    \label{fig: fmodvszshell1}
\end{figure}

\begin{figure}[htbp]
    
    \begin{subfigure}[b]{0.48\textwidth}
        \centering
        \includegraphics[width=\linewidth]{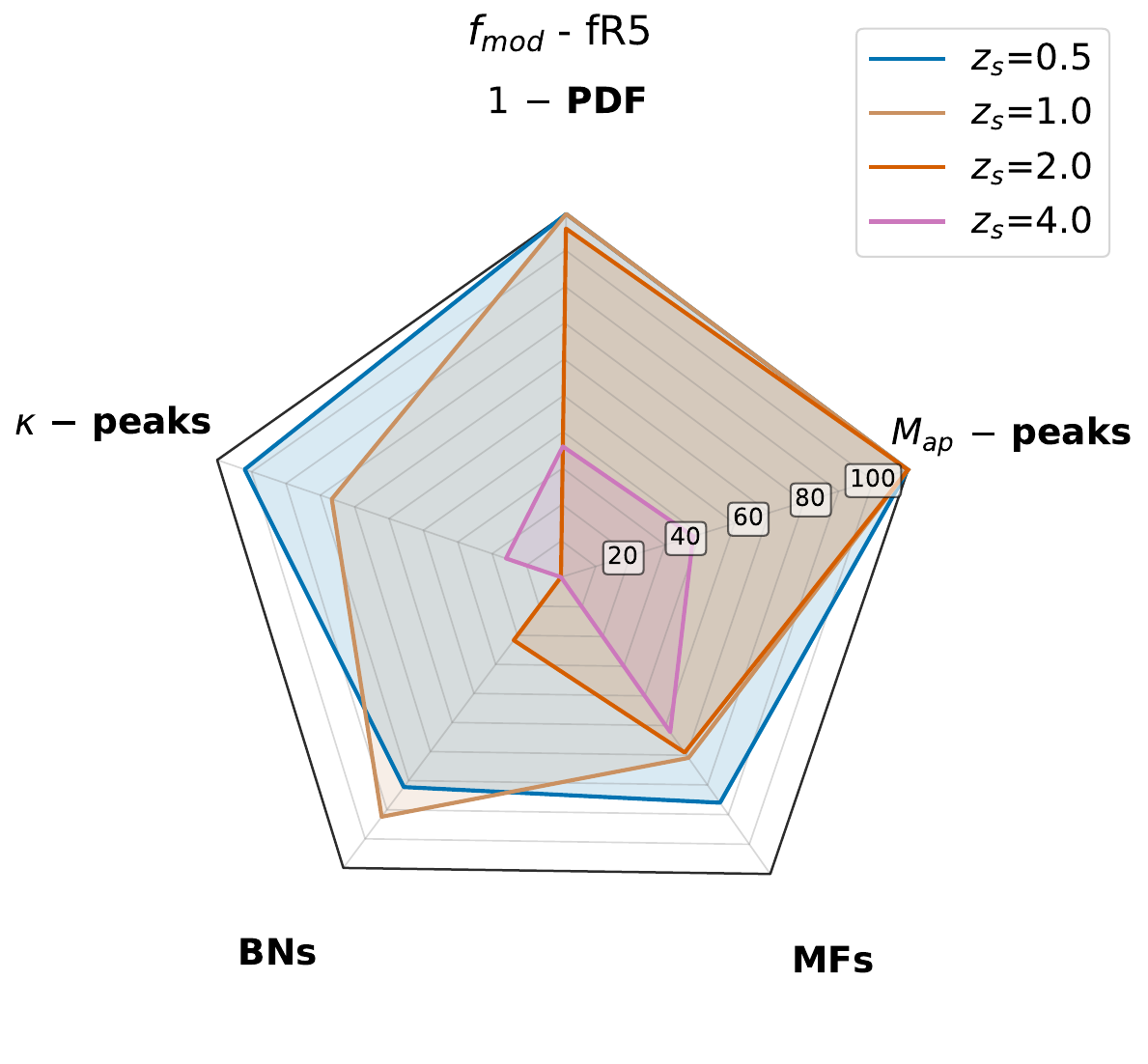}
    \end{subfigure}
    \begin{subfigure}[b]{0.48\textwidth}
        \centering
        \includegraphics[width=\linewidth]{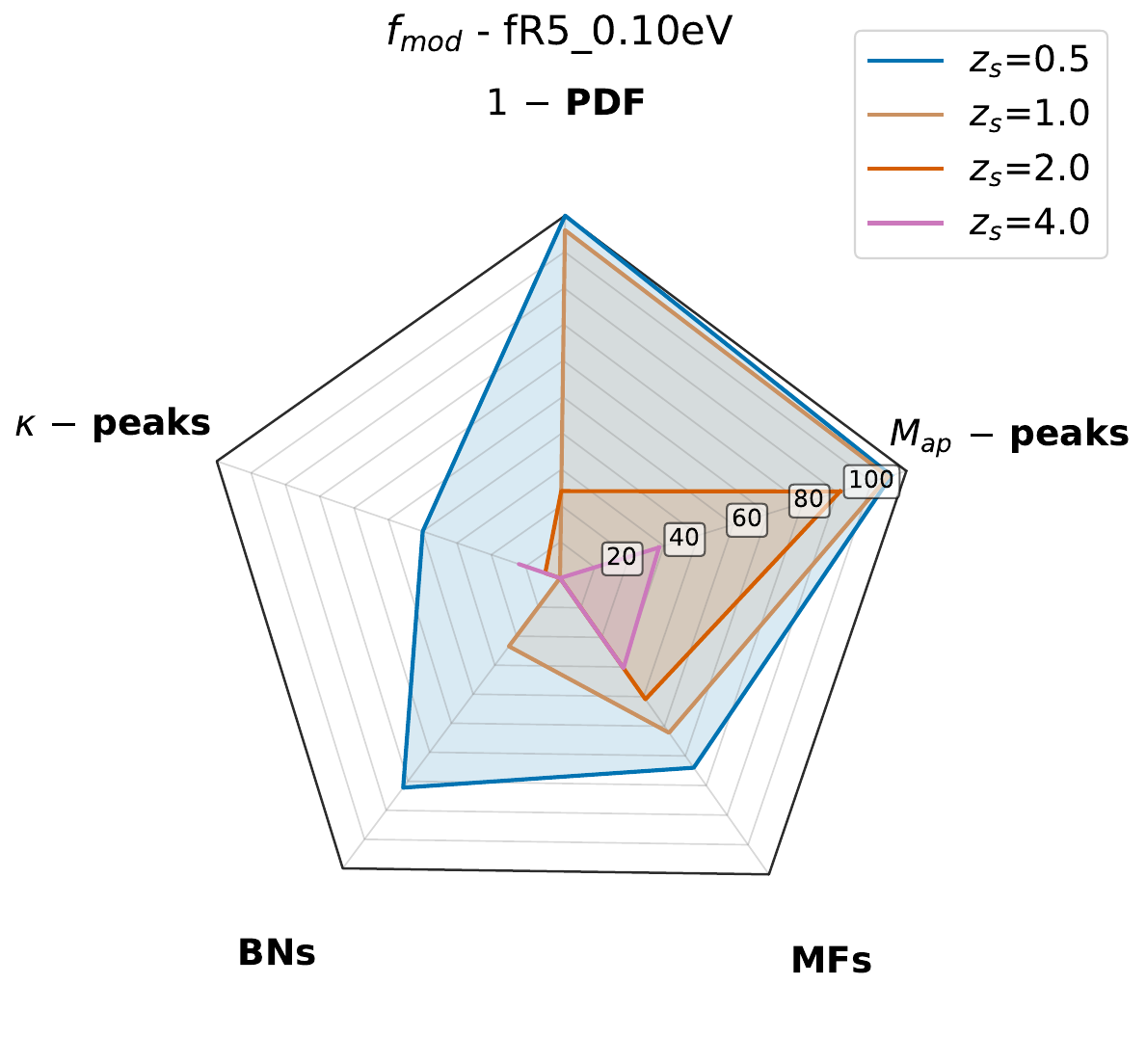}
    \end{subfigure}
    

    \begin{subfigure}[b]{0.48\textwidth}
        \centering
        \includegraphics[width=\linewidth]{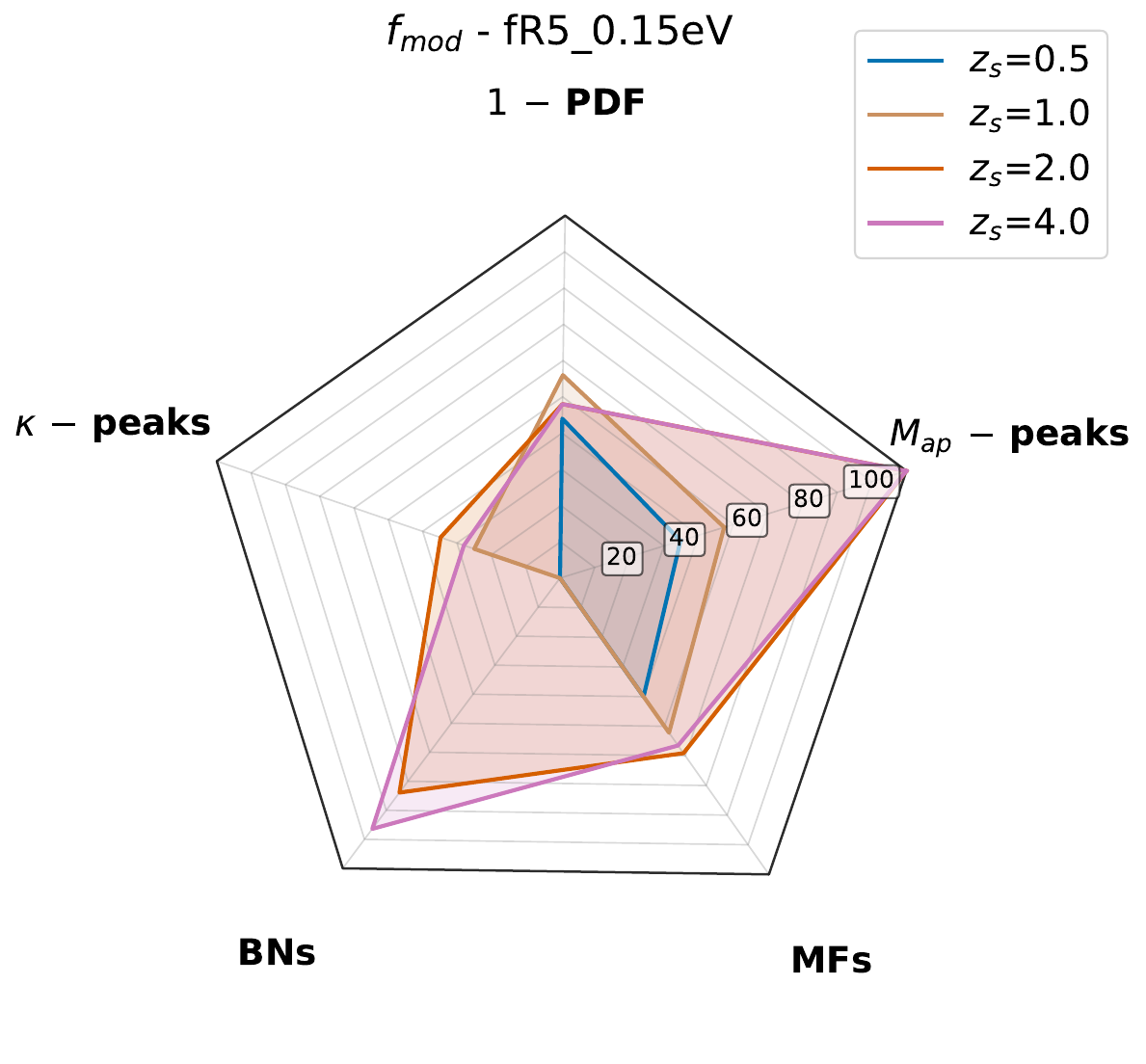}
    \end{subfigure}
    \begin{subfigure}[b]{0.48\textwidth}
        \centering
        \includegraphics[width=\linewidth]{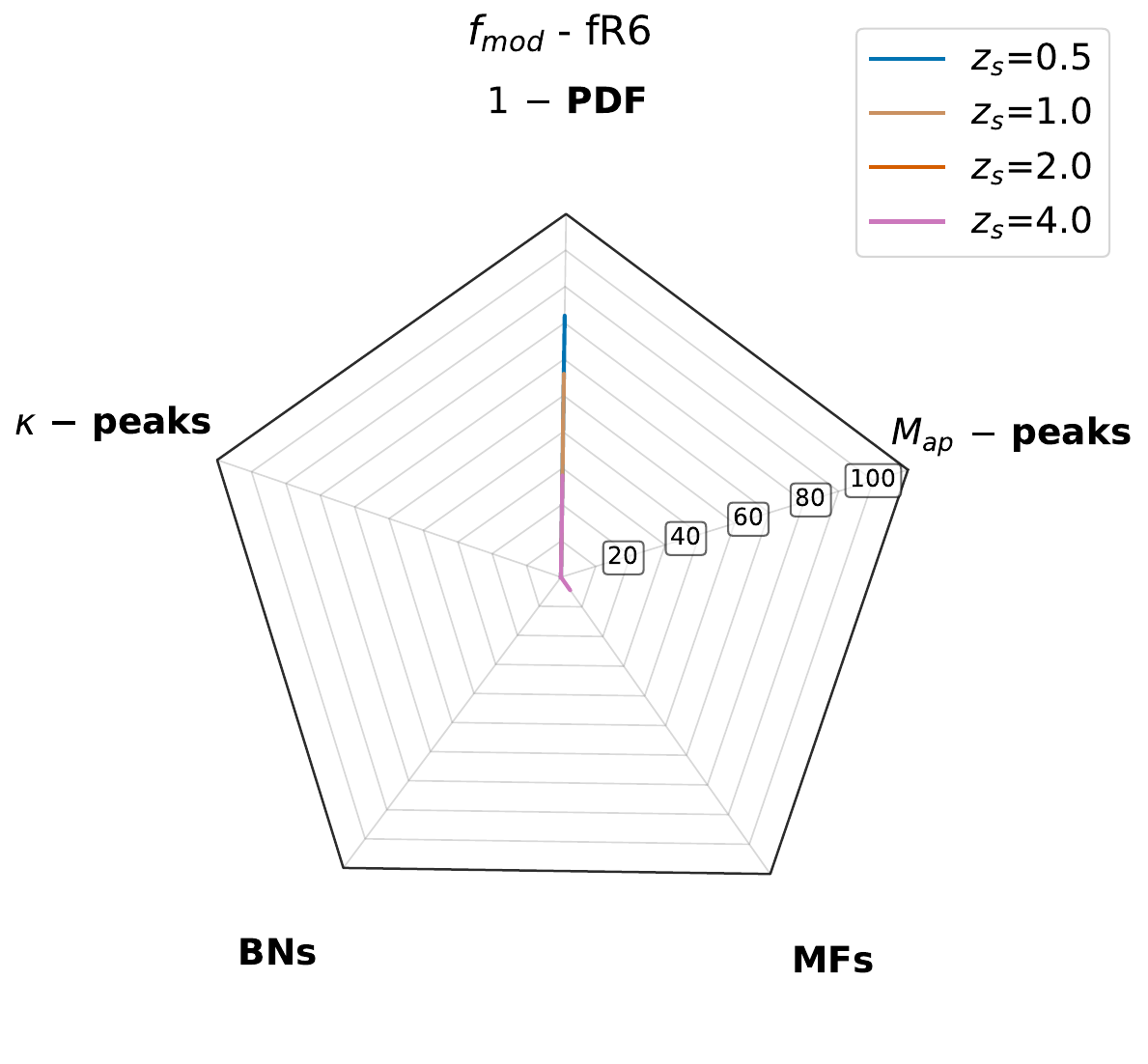}
    \end{subfigure}
    

    \begin{subfigure}[b]{0.48\textwidth}
        \centering
        \includegraphics[width=\linewidth]{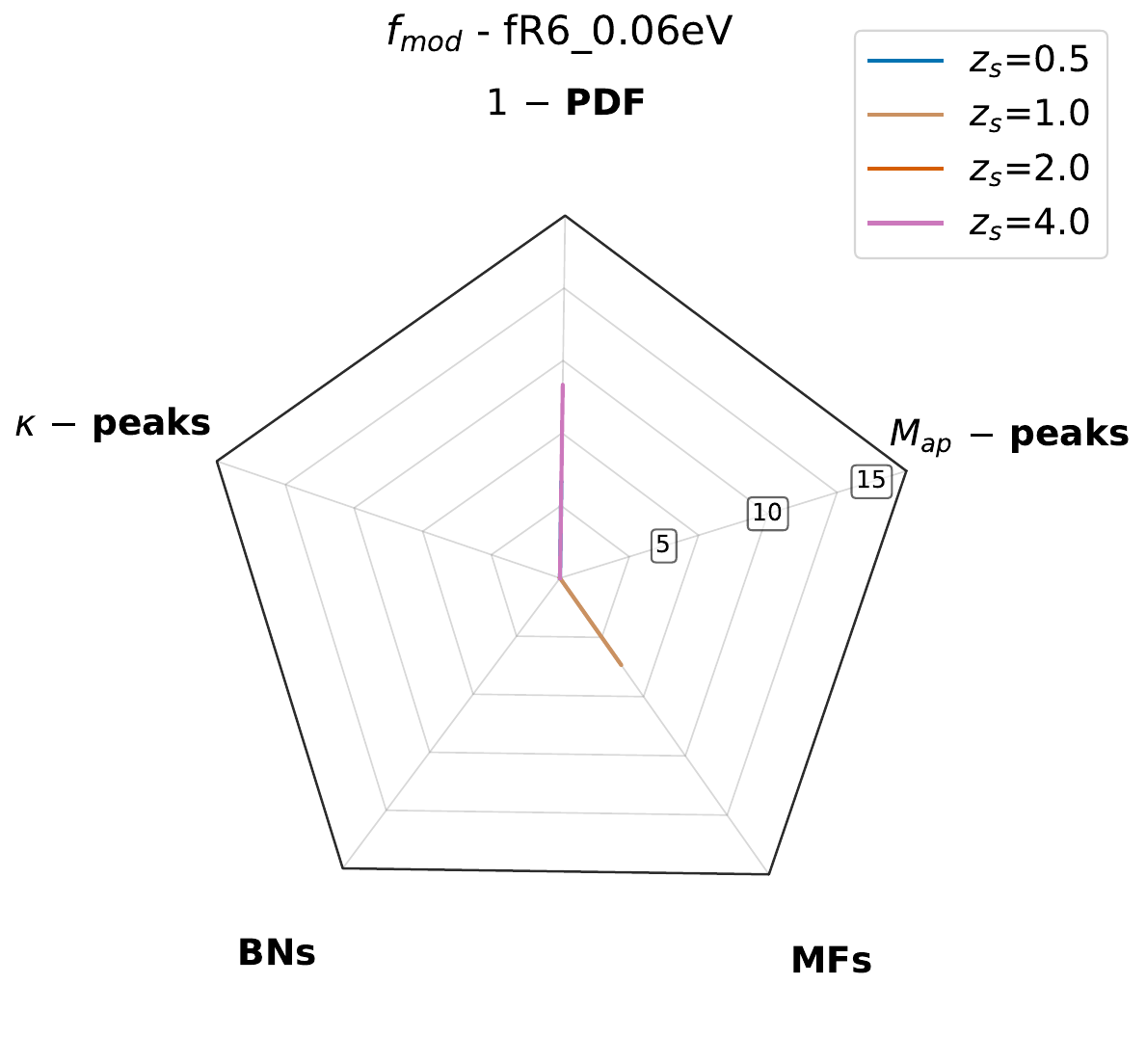}
    \end{subfigure}
    \begin{subfigure}[b]{0.48\textwidth}
        \centering
        \includegraphics[width=\linewidth]{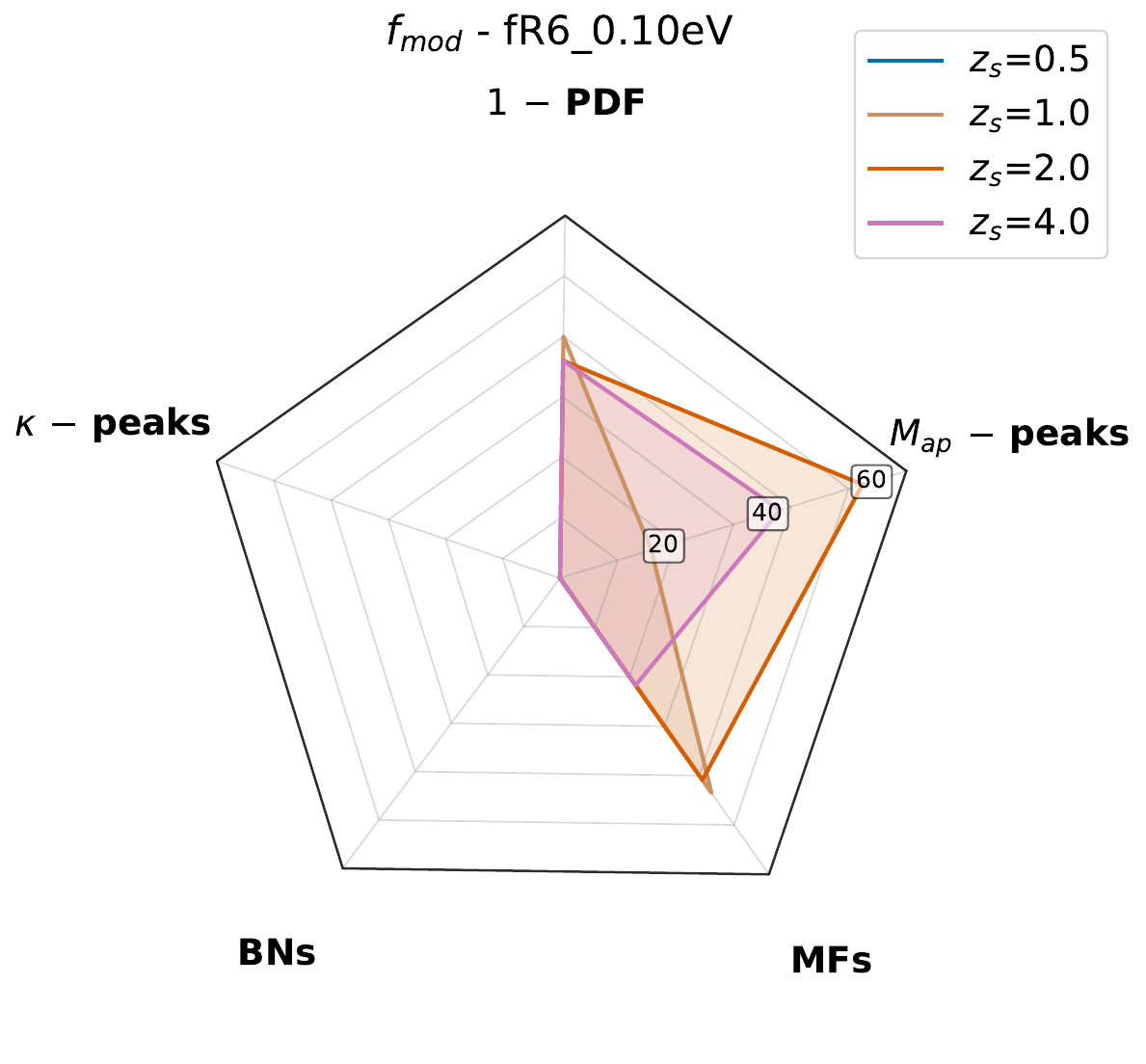}
    \end{subfigure}
    
    \caption{Same as Fig.\,\ref{fig: fmodvszshell1} for fR5 and fR6 cosmologies. Please note the scale change in the last two fR6 plots for better viewing.}
    \label{fig: fmodvszshell2}
\end{figure}

In order to help the reader to easily compare the performances of the different HOS probes, we use spider\,-\,web plots as the one in Fig.\,\ref{fig: fmodvszshell1}. Here, each point gives the value of the test for the probe labeled at the corresponding vertex of the spider\,-\,web, the distance from the centre being larger for larger values. In such a plot, the more the sides of the polygon obtained connecting the points are parallel to the spider\,-\,web border, the more the different probes perform in a similar way. 

Fig.\,\ref{fig: fmodvszshell1} helps understanding how to read these spider\,-\,web plots. In the left panel, we show the results for the fR4 model, each line referring to a different source redshift. All the lines are close to the border, and almost parallel to it indicating that $f_{mod}({\cal{O}}, {\cal{M}})$, setting ${\cal{M}} = d_H$, almost saturates to its maximum value for all the the five HOS probes ${\cal{O}}$ considered. This is the same as saying that all the probes can easily discriminate between $\Lambda$CDM and fR4 model at all redshift. This is not surprising given that this particular case is the one with the largest (and actually unrealistic) value of $\log{|f_{R0}|}$ hence leading to the strongest departure from GR. It is nevertheless interesting to compare the results in the left panel to those in the right one, which refers to the case with the same $\log{|f_{R0}|}$ but with $M_{\nu} = 0.3 \ {\rm eV}$ massive neutrinos. Most of the lines now collapse towards the centre, i.e., towards smaller values of $f_{mod}({\cal{O}}, {\cal{M}})$. A residual discriminatory power, i.e., values of $f_{mod}({\cal{O}}, {\cal{M}}) > 0$ at all $z_s$, is obtained for the 1\,-\,PDF, and the MFs, while peaks and BNs are only marginally useful at the largest redshifts. This is a textbook example of how massive neutrinos can compensate for the enhanced clustering of $f(R)$ theories, leading to an almost perfect cancellation. As a consequence, deviations from GR become very difficult to detect, and HOS retain meaningful discriminatory power only in a limited set of cases.

Both the values of $\log{|f_{R0}|}$ and $M_{\nu}$ in the previous cases are extreme, and mainly reported for illustrative purposes. Nevertheless, the effect is fully general. For each value of $\log{|f_{R0}|}$, it is possible to find a neutrino mass which compensates the effect of MG hence leading to a significant drop of $f_{mod}({\cal{O}}, {\cal{M}})$ for all HOS probes. On the other hand, one should also be careful to not overcompensate. Indeed, should the value of $M_{\nu}$ be too large, one can again found deviations from the $\Lambda$CDM reference case, but due to the suppressed clustering caused by neutrinos rather than the enhanced one due to $f(R)$. This can be seen looking at the results in Fig.\,\ref{fig: fmodvszshell2} for the fR5, fR5$\_$0.10eV, and fR5$\_$0.15eV models. In the case of massless neutrinos, the deviations from GR due to the $f(R)$ with $\log{|f_{R0}|} = -5$ are still large enough that most of the HOS probes considered can detect them at small and intermediate $z_s$. When massive neutrinos are added, the scaling of $f_{mod}$ depends on which of the two contrasting effects, i.e., the enhanced or suppressed clustering due to $f(R)$ and massive neutrinos, dominate. A complicate interplay takes indeed place. First, there is a suppression for $M_{\nu} = 0.10 \ {\rm eV}$ with the effect being more noticeable at large $z_s$ because here the deviations from GR are less important. Increasing the mass to $M_{\nu} = 0.15 \ {\rm eV}$ suppresses the $f_{mod}$ values at low $z_s$, but makes them larger at high $z_s$. This is because the model now deviates from the fiducial $\Lambda$CDM one, not due to deviation from GR, but for the impact of massive neutrinos compared to the massless case taken as reference. This is still more evident for the cases with $\log{|f_{R0}|} = -6$ since one is now actually investigating the impact of the neutrino mass in a quasi\,-\,GR theory. The larger $M_{\nu}$ then, the larger $f_{mod}$ for all HOS probes, whatever is the $z_s$ value.

The coarse sampling in $z_s$ is the cause of the noisy trends of $f_{mod}$ with the source redshift, but it is not responsible for the fact that there is no monotonic behavior valid for all models and probes. Indeed, the scaling with $z_s$ is a complicated interplay among the strength of the deviations from GR, the impact of neutrino mass, and the HOS probe S/N (given that the covariance enters the definition of $d_H$ hence determining the significant of the MG signature). Since the covariance matrix enters the computation of the Hellinger distance, measurements at low $z_s$ should be preferred since the data are less affected by noise, hence the S/N is larger. When this is the dominant driver, we indeed observe a decreasing trend of $f_{mod}({\cal{O}}, {\cal{M}})$ with $z_s$. On the contrary, if the driving effect is the growth suppression due to a neutrino mass so large to not be compensated by $f(R)$ enhanced clustering, we get a larger discriminatory power at high $z_s$. Note that different trends would have been observed had we chosen a different metric, since the dependence on the S/N would have been different. For instance, the weighted and unweighted SMAPE (${\cal{S}}_{NW}, {\cal{S}}_{YW})$, and the Wasserstein distance $d_W$ do not include the covariance in their definition, so that how the S/N scale with $z_s$ is not important.

Finally, it is worth wondering whether there is a HOS probe which works better in terms of discriminatory power. To this end, one could look at Figs.\,\ref{fig: fmodvszshell1}, \ref{fig: fmodvszshell2} and search for the probe with the larger $f_{mod}({\cal{O}}, {\cal{M}})$. Such an analysis, however, would not give a unique answer since which is the best HOS probe depends on both the model and the source redshift. For instance, at $z_s = 0.5$, the 1\,-\,PDF and the $M_{\mathrm{ap}}$\,-\,peaks provide larger $f_{mod}$ values, while the topological ones, either BNs or MFs, are less efficient. However, MFs recover discriminatory power at $z_s = 4.0$ becoming comparable if not better than the convergence PDF or the aperture mass peaks. The unique definitive conclusion is that, among the two peaks statistics, the one based on the convergence map is way less constraining than the one obtained after computing the aperture mass. It is also worth noticing that the comparison among probes is also complicated by the fact that both BNs and MFs data vectors are actually made out of different elements related to different higher orders. For instance, MFs are made out of $(V_0, V_1, V_2)$ with $V_0$ being a 2nd order probe. Should this term dominate the overall S/N, the MFs actually reduces to a quasi\,-\,2nd order statistic hence decreasing their discriminatory power. On the contrary, the 1\,-\,PDF and the $M_{\mathrm{ap}}$\,-\,peaks are always sensitive to the full high order expansion, hence retaining a larger discriminatory power provided the S/N is large enough (which is the case at low $z_s$). However, the lack of a unique probe standing out over the others should be considered a good news. Indeed, no matter the model or the source redshift, we can safely argue that it is always possible to find a HOS probe having a discriminatory power sufficient to distinguish among GR and $f(R)$ theories.

\section{Results from hypothesis testing method}
\label{sec:nonparametric_results}

Metrics-based methods proved to be reliable in assessing the discriminating power of the HOS, although the Gaussianity requirement of the likelihood and the error assignment procedure, as we mentioned in Sec.\,\ref{sec:metrics}, forced us to exclude HOM from the analysis. Nonparametric methods, on the other hand, not relying on any assumption on the likelihood, allow us to compare also the performance of the HOM with respect to the other HOS. They should be then intended as a complementary method with respect to the metrics to let us consider the whole set of HOS. It should be noted that, with respect to the metrics that provide more extensive information on the whole behavior of the selected probe, nonparametric tests instead give an estimate of the point-like maximum distance, reached along the graph of the probe, between the GR and MG models.

After this clarification, let us now discuss the discriminatory power of the different HOS as evaluated through the hypothesis testing methods introduced above. We have carried on this analysis, varying for each model the configuration setup of the HOS probe, and for different probes. We also remind the reader that, for each choice, we have 20 histograms, one for each ${\cal{O}}$ bin, so that the number of combinations is so large that showing all of them would make this paper unacceptably long. Moreover, a caveat is in order here. Let us suppose that we find a $p_i \le p_{th}$ for the i\,-\,bin of a given HOS for a chosen configuration. This does not imply that $p_i < p_{th}$ for all bins. This is because some bins are actually not informative. For instance, the extreme tails of the 1\,-\,PDF or of Betti numbers go to zero for all models just because of their same definitions, so these bins can not be used to discriminate among different models. What matters for our aims is, however, that we find evidence for statistically meaningful deviations in at least one bin. One could argue that the better probes would be those where the number of bins with $p < p_{th}$ is larger, but this comparison would be unfair since it would not take into account the peculiarities of each probe. Motivated by this consideration and by the need to avoid cluttering the paper with a list of figures, we will therefore only report here some illustrative cases showing the most relevant bins to support the general results found.

\subsection{Choosing the hypothesis testing method}

We first want to investigate which of the four options introduced in Sec.\,\ref{subsec: nonparamstests} is best suited to pinpoint the differences among GR and $f(R)$ models. We compute these tests for all bins in all possible combinations, both on the raw samples and on the smoothed distributions by KDE. We noticed that the use of KDE led to better discrimination and, as an additional benefit, helped us to discard the uninformative bins (like, e.g., the extreme bins of the 1\,-\,PDF, which are all null in every cosmology). This should not be a surprise, since most of the statistical tests used assume continuously distributed data. 

As a preliminary remark, we remind the reader that we have more than one bin for all HOS probes, but the HOM that are measured at a single value for each realization. We therefore decided to make the comparison among the different methods and probes using only the bin with the largest discriminatory power. Note also that, for 2PCF, we do not perform any smoothing before the measurement, so that the results will be the same when comparing with other probes at varying smoothing.

\begin{figure}[h]
    \centering 

    \begin{minipage}[b]{0.48\textwidth}
        \includegraphics[width=\textwidth,keepaspectratio]{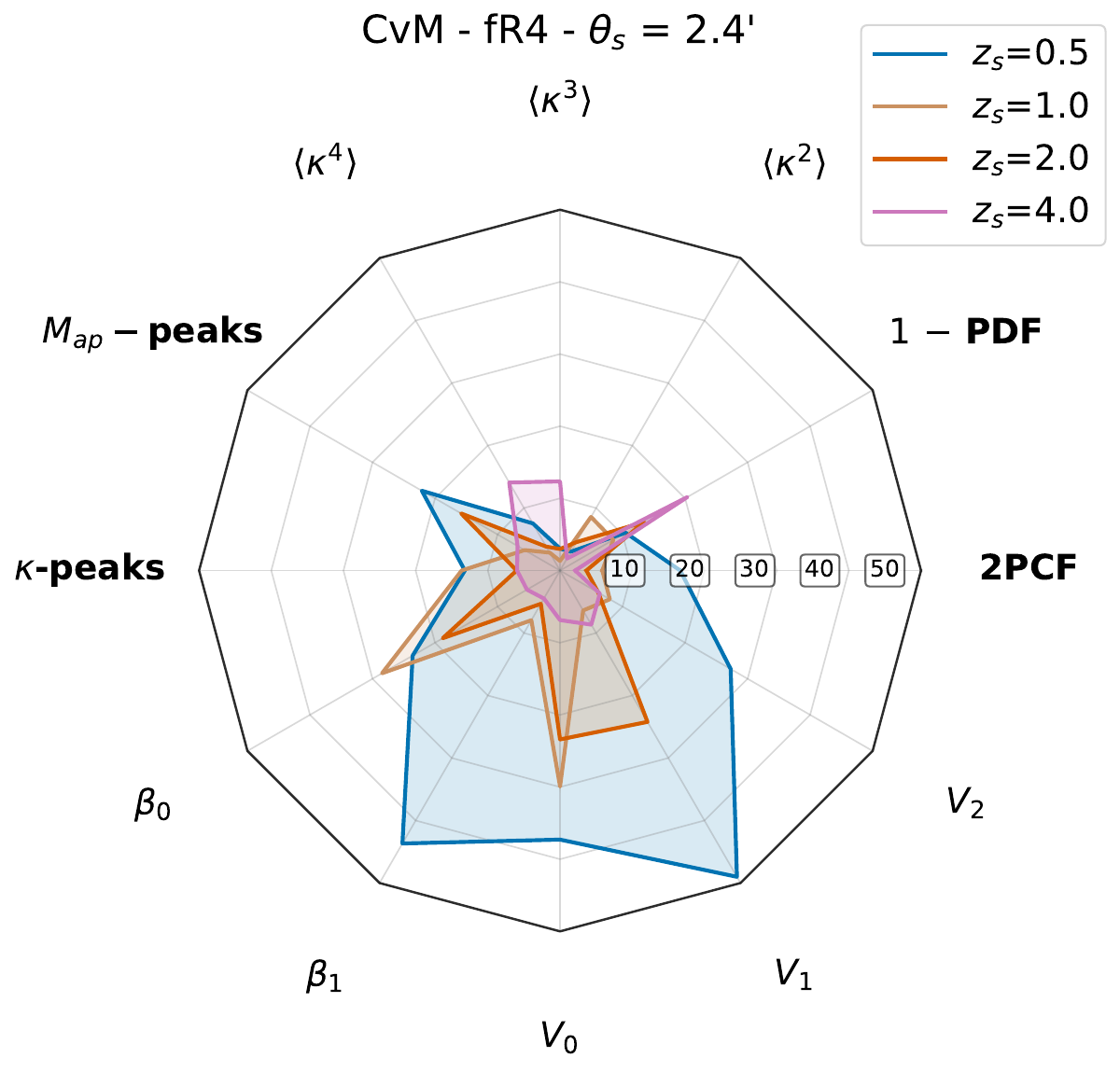}
    \end{minipage}
    \hfill 
    \begin{minipage}[b]{0.48\textwidth} 
        \includegraphics[width=\textwidth,keepaspectratio]{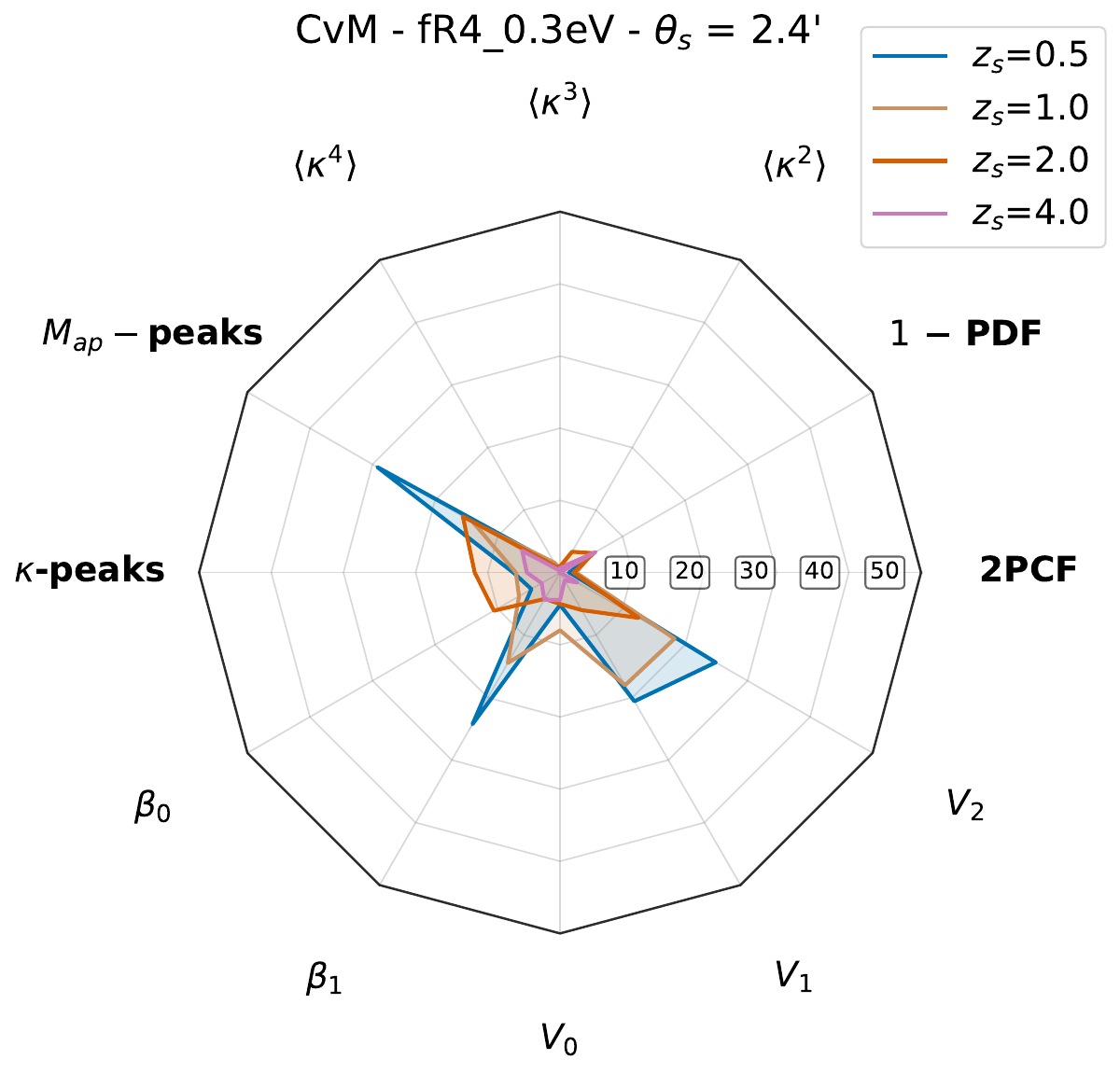}
    \end{minipage}

    \caption{CvM test results for the fR4 model setting $M_{\nu} = 0$ (left) or $M_{\nu} = 0.3 \ {\rm eV}$. Here, and in the following plots, we set $\theta_s = 2.4'$, while the different colors refer to the four $z_s$ values.}
    \label{fig:cvm_fr4_mv_grids}
\end{figure}

We noticed that the AB test does not provide a clear differentiation of the HOSs, its value being quite homogeneous among the different probes. The AB test indeed evaluates the ratio of the variance of the samples and, given the strong overlap hence the similar spread of the distributions, the AB test can hardly identify great differences between different probes. A similar argument also applies to the MWU test, showing the same trend for different configurations. We will then refrain from showing other results regarding these tests, focusing on the ones providing clearer results. In Fig.\,\ref{fig:cvm_fr4_mv_grids}, we can see indeed how the CvM test works in differentiating the HOS performances in the case of fR4 cosmologies for the same model and setup as checked for the AB one. It is evident that the CvM test performs better with respect both to AB and MWU ones at pointing out which HOS probe has a larger discriminatory power. Similar trends are also found for the KS test, although less accentuated.

The better performances with respect to the AB and MWU cases are due to the fact that both KS and CvM tests are more sensitive to the distance between the distributions, which, as we saw in previous paragraph, can still be present even when the histograms overlap and have a similar spread. On the other hand, the reason why CvM performs better than KS is related to the fact that they evaluate the differences between samples in different ways. On one hand, KS computes the largest absolute vertical distance between the two empirical cumulative distribution functions of the samples, in this way being particularly sensitive to large, isolated deviations between the distributions. In contrast, CvM is based on the integrated squared difference between the two empirical distribution functions, providing a more comprehensive overview of the discrepancy across the entire range of the data, measuring cumulative deviations. Since it turns out that GR and MG samples for each HOS probe show more global differences with respect to single point large deviations, the CvM emerges as the most suitable test to distinguish them. For these reasons, in the following, we will rely on the CvM test only to investigate how different HOSs are efficient at discriminating among GR and $f(R)$ models.

\subsection{Discriminating models}

Let us now look at the results for the different models of interest. We will rely on the CvM test, considering only the case giving the strongest discrimination. We will first fix the smoothing at $\theta_s = 2.4 \ {\rm arcmin}$ since we find this choice provides better results.

Fig.\,\ref{fig:cvm_fr4_mv_grids} shows the results for the fR4 models. In the left panel, we consider the case without massive neutrinos. This is the case, giving the strongest deviation from GR, so it is not surprising that all of the probes are quite efficient at discriminating it. It is nevertheless worth noticing that, even in this favorable case, the HOS probes, for a fixed $z_s$, provide larger values than the 2PCF, thus confirming the advantage of going beyond second-order statistics. The strongest constraining power is given by the topological probes $(V_0, V_1, \beta_1)$ at $z_s = 0.5$, while the ranking is different at larger $z_s$ with $(1\,-\,{\rm PDF}, \langle \kappa^3 \rangle, \langle \kappa^4 \rangle)$ being the preferred choice at $z_s = 4$. A dramatic drop in the discriminatory power is observed (in the right panel) if we allow the neutrino to have a mass $M_{\nu} = 0.3 {\rm eV}$, a value which is actually quite unrealistic given the most recent constraints on $M_{\nu}$. Such a large value is needed to compensate for the large boost in clustering due to $f(R)$ with large $f_{R0}$, and is shown here more as a pedagogical case illustrating the opposite effects of $f(R)$ and massive neutrinos. We indeed find that now all the probes become less able to discriminate between GR and MG. It is worth noticing, however, that the MFs $(V_1, V_2)$ measured at $z_s = 0.5$ still retain some constraining power, while the $M_{\mathrm{ap}}$\,-\,peaks are the least affected probe with the CvM value at all $z_s$ being almost the same as in the massless case.

\begin{figure}[h]
    \centering

    \newcommand{\slotwidth}{0.48\textwidth} 
    \begin{subfigure}[t]{\slotwidth}
        \centering
        \includegraphics[width=\linewidth, keepaspectratio=false]{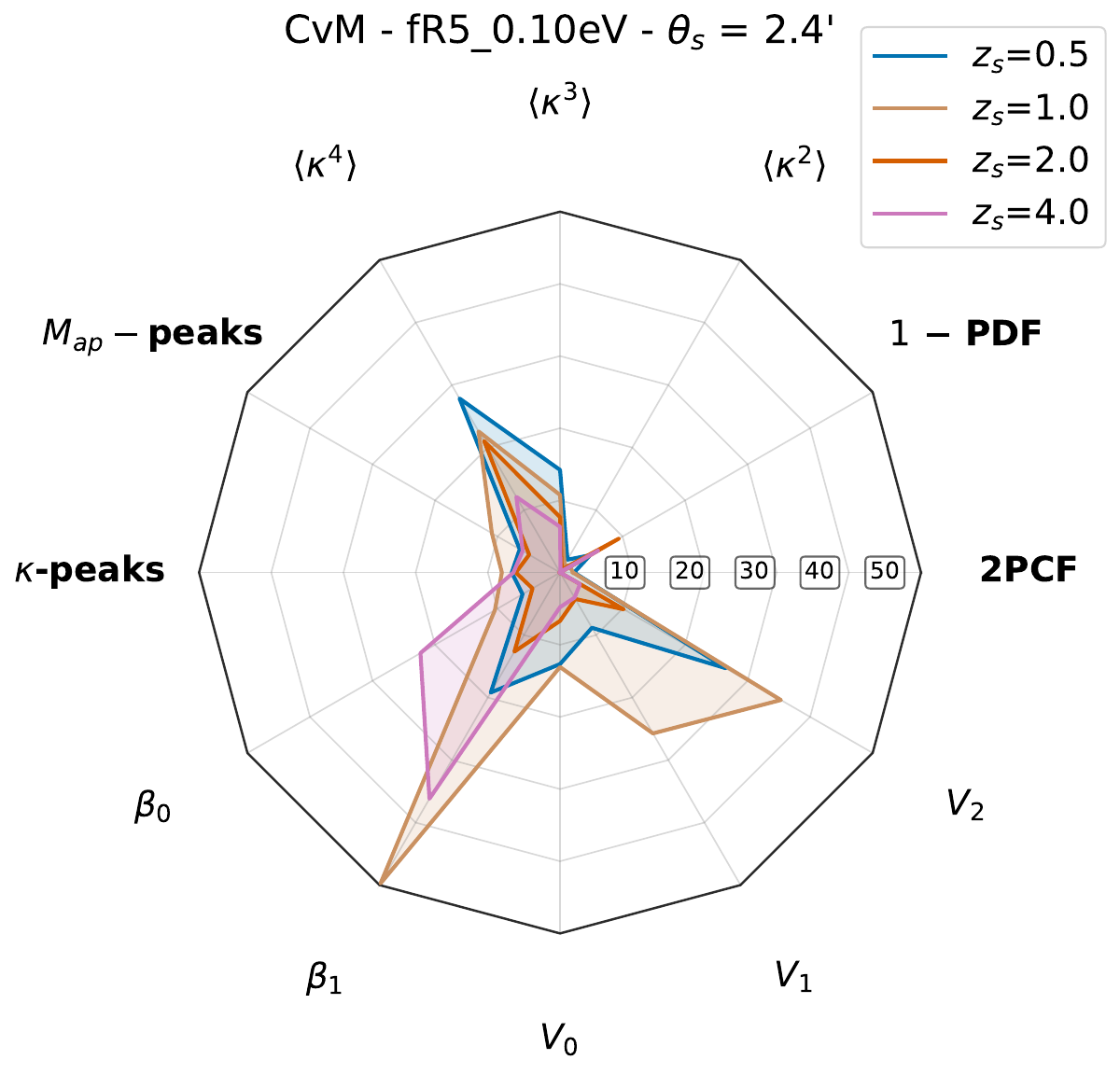}
    \end{subfigure}%
    \hfill%
    \begin{subfigure}[t]{\slotwidth}
        \centering
        \includegraphics[width=\linewidth, keepaspectratio=false]{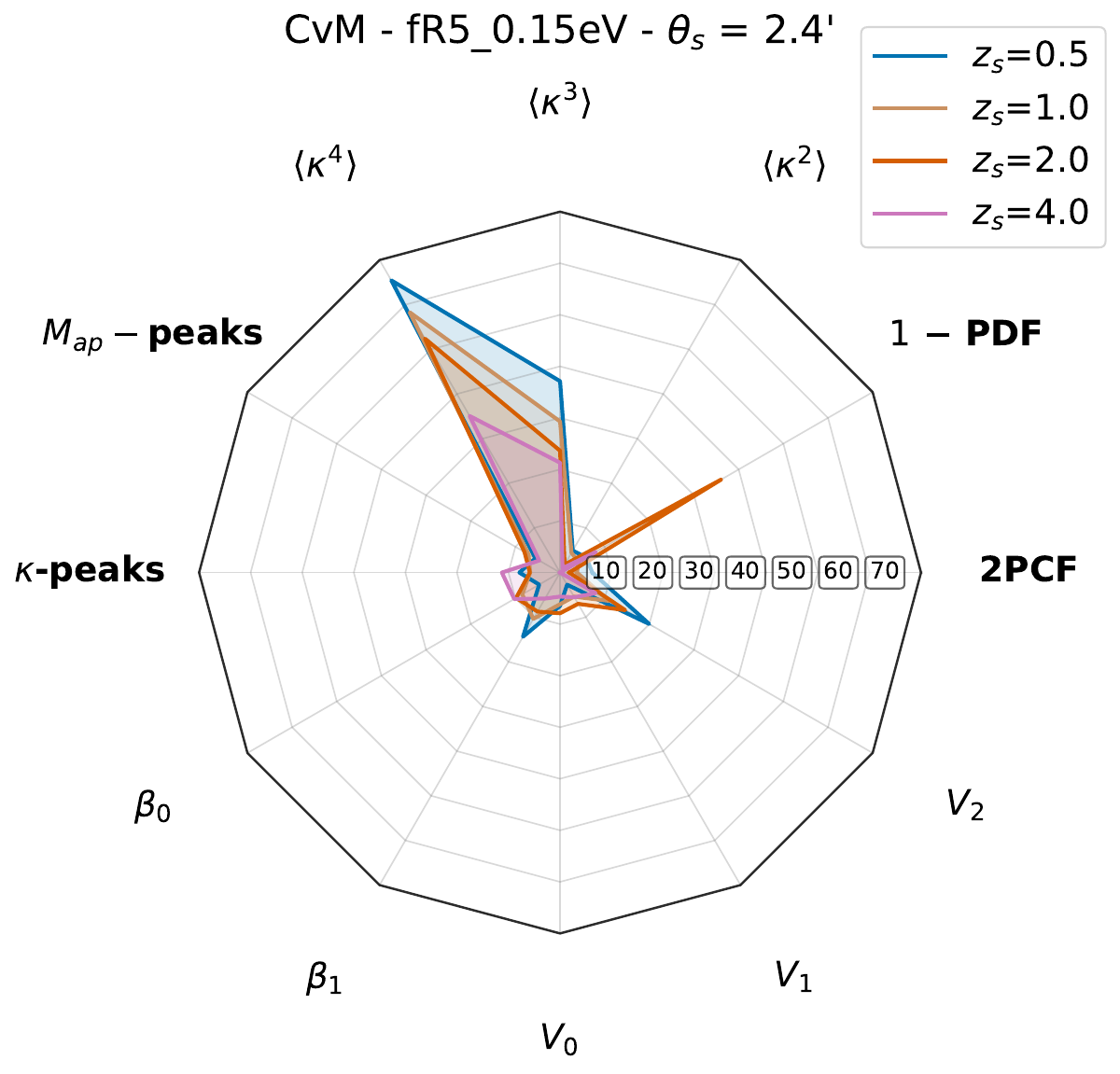}
    \end{subfigure}
    \caption{Same as Fig.\,\ref{fig:cvm_fr4_mv_grids} but for the fR5 models with $M_{\nu} = (0.10, 0.15) \ {\rm eV}$ respectively in the left and right panels. Beware of the different scale used in the right spider\,-\,web plot.}
    \label{fig:cvm_fr5_mv_grids}
\end{figure}

Moving to fR5, we see in Fig.\,\ref{fig:cvm_fr5_mv_grids} the comparison for the two cases with $M_{\nu} = 0.1$eV and $M_{\nu}=0.15$eV. We can see in general that the discriminating power degrades with increasing neutrino masses, although there are exceptions. The PDF is, in fact, affected at lower redshifts, while the HOMs, in particular $\langle \kappa^4 \rangle$, tend to perform very well at higher redshifts also with neutrinos. Topological statistics as well, like $\beta_1$, tend to provide a very good discrimination. A general trend is that almost every HOS outperform two-point statistics in discriminating power; both $\langle \kappa^2 \rangle$ and the 2PCF obtain very low values of the CvM statistic, and in the case of $M_{\nu}=0.1$eV, represented in left panel of Fig.\,\ref{fig:cvm_fr5_mv_grids}, the 2PCF does not even show a statistically significant difference between GR and MG samples.

\begin{figure}[h]
    \centering

    \newcommand{\slotwidth}{0.48\textwidth} 
    \begin{subfigure}[t]{\slotwidth}
        \centering
        \includegraphics[width=\linewidth, keepaspectratio=false]{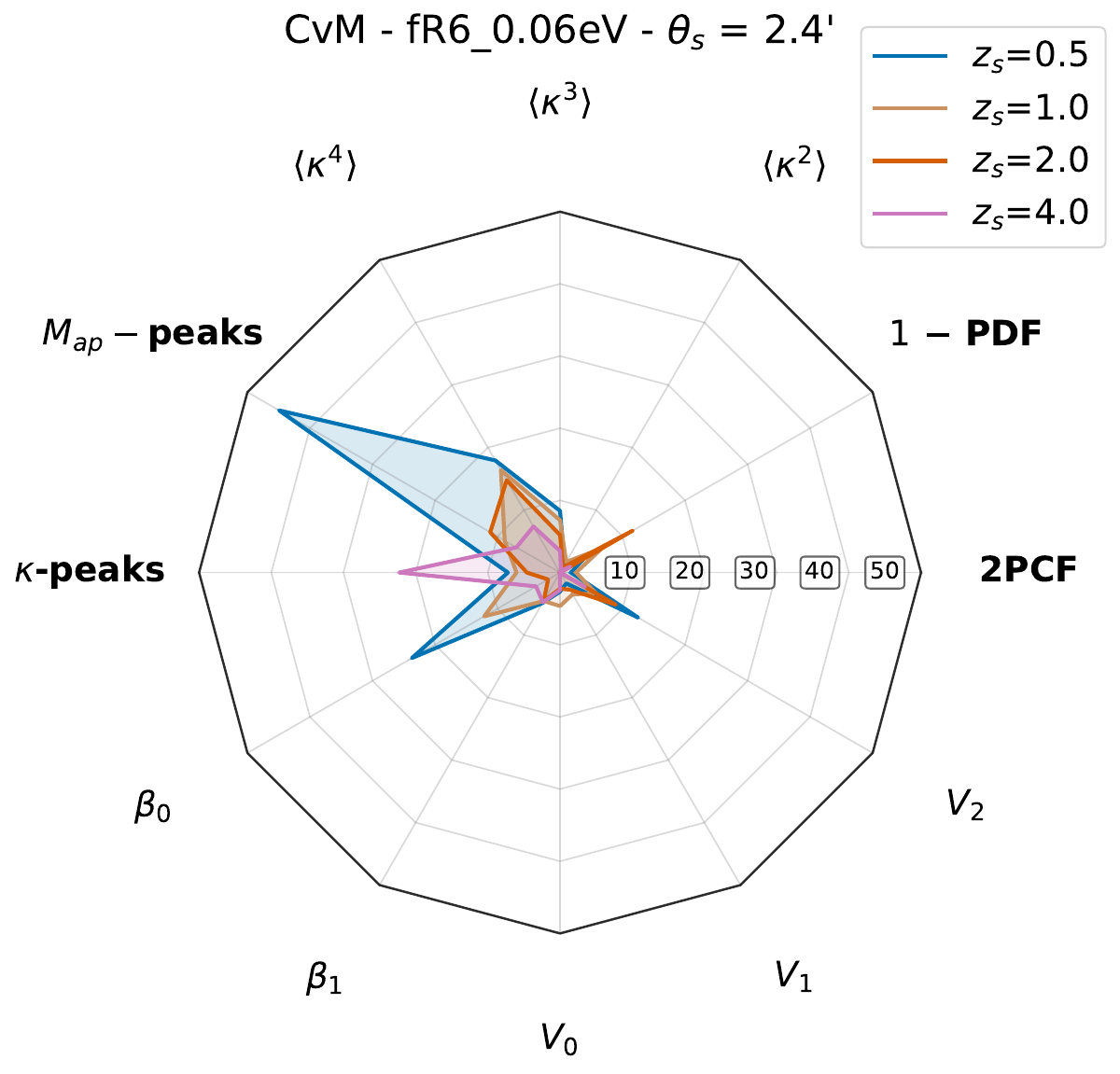}
    \end{subfigure}%
    \begin{subfigure}[t]{\slotwidth}
        \centering
        \includegraphics[width=\linewidth, keepaspectratio=false]{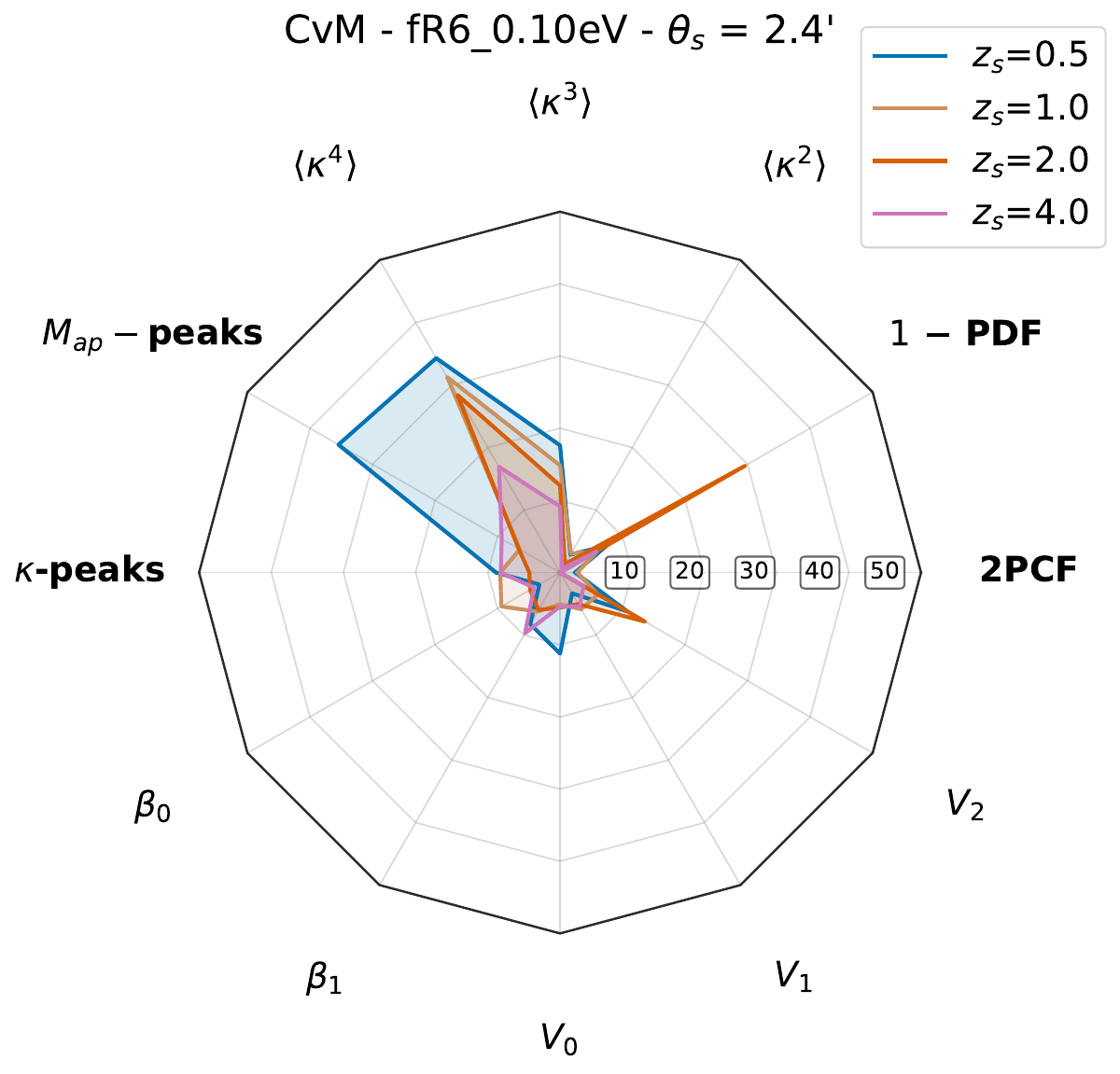}
    \end{subfigure}
    \caption{Effect of massive neutrinos, $M_{\nu} = (0.06, 0.1) $eV, on the discriminating power for different HOSs according to the CvM test, for the fR6 cosmologies, at fixed smoothing scale.}
    \label{fig:cvm_fr6_mv_grids}
\end{figure}

Finally, we consider the weakest modifications of gravity in our models, i.e., the fR6 cosmologies. We can see how the HOSs perform according to CvM test, when adding massive neutrinos, in Fig.\,\ref{fig:cvm_fr6_mv_grids}. As happened in previous cases, between the weak MG signature and the effect of noise, we do not see a clear pattern among the different cases, although it is evident that HOS always outperform two-point statistics. As we saw for the fR5 cosmologies HOM, and specifically $\langle \kappa^4 \rangle$, tend to have a good discriminatory power. However, for the fR6\_0.06eV cosmology, the one closer to $\Lambda$CDM, we can see how $M_{\mathrm{ap}}$\,-\,peaks and 1\,-\,PDF are the ones that allow us to better distinguish between models.

\section{Conclusions}
\label{sec:conclusion}

The possibility to probe in a single step both visible and dark matter made gravitational lensing an attractive tool for investigating the properties of this dominant component of galaxies and clusters of galaxies. It was then soon realized that, in the weak regime, gravitational lensing could also be used to probe both the background expansion and the growth of structures, thus pointing at cosmic shear as one of the best weapons in the panoply of techniques Stage III surveys could use to investigate dark energy. With the advent of Stage IV surveys, it is now possible to make a step further in two directions. On one hand, one can think of discriminating between dark energy and modified gravity. On the other hand, this same goal can be achieved by going beyond the standard second-order probes, being the next\,-\,to\,-\,come data of quality and quantity sufficient to measure higher-order statistics. Recent results based on the final release of Stage III surveys have already confirmed that this promise can be fulfilled \cite{martinet2018, gatti_des2022, harnois_deraps2024}, while what Euclid can do has been convincingly shown in the analysis carried out by the HOWLS team \citep{euclid_howls2023,euclid_howls2025}.

Motivated by this scenario, we have here carried out a critical analysis of the use of HOS as a tool to discriminate among GR and MG, looking at the {\it what, which, how} aspects of this issue. To this end, we have relied on the DUSTGRAIN - \textit{pathfinder} suite of simulations to create noisy convergence maps to be given as input to the same publicly available codes that we have used as members of the HOWLS team to measure a wide set of HOS probes. The cosmological parameters were fixed on Planck 2015 cosmology, varying  $\log|f_{R0}|$ and $M_{\nu}$ for 8 different parameters combinations. We had 256 light-cone realizations, from which we obtained square convergence maps of $5\times5$ deg$^2$. We injected a Gaussian distributed shape noise with ellipticity dispersion $\sigma_\epsilon = 0.26$ and galaxy number density $n_{\rm gal}=30\,\mathrm{arcmin}^{-2}$. In this first step of a larger project, we have considered all sources to be at the same redshift $z_s$, choosing four different values. Although simplistic, such an assumption avoids the need to choose a redshift distribution so that the results are not survey-specific, but rather tell us what is the redshift where the signatures of MG are most evident.

We have first investigated {\it what} kind of MG theories we can discriminate. The Hu\,-\,Sawicki $f(R)$ model has been chosen as a prototype example of the enhanced growth of structures the MG can produce. However, one should also worry about elements that could possibly contrast this effect so that the final result is a MG model looking the same as a GR one. This is indeed the case if massive neutrinos are included since they suppress the growth thus possibly washing out the signature of $f(R)$ model. This is exactly what we find. Even in the case of large deviations from GR, as for $\log{|f_{R0}|} = -4.0$, it is possible to tailor the neutrino mass so that the discriminatory power of any statistics (2nd order or HOS) is strongly reduced, if not removed at all. It is, however, worth noticing that the neutrino mass needed to achieve this result is sometimes too large to be realistic, at least under the assumption of $\Lambda$CDM. Moreover, its value should be somewhat finely tuned since a too large one make the deviations from GR emerge again. It should also be considered that the intertwining between MG and massive neutrinos can be more convoluted. Current cosmological bounds on neutrino masses are derived under the GR assumption, and the seemingly unrealistic values of $M_{\nu}$ within this framework could become viable once GR is relaxed in favor of modified models. In addition, the overcompensation of MG effects by neutrinos could also come from departures from the minimal-neutrino-mass even within $\Lambda$CDM, again leading to a non-detection of MG signatures, but for different reasons. For a fully consistent interpretation of the results, all the effects mentioned above should be taken into account. An extensive analysis more finely sampling the $\log{|f_{R0}|}\,-\,M_{\nu}$ parameter space, is needed to understand when a signature of deviation from GR can be attributed to $f(R)$ or massive neutrinos looking at the redshift regimes where their two opposite effects dominate.

Most of the previous works relying on HOS have focused on a single probe, such as high order moments \citep{gatti_des2022}, peaks count \citep{martinet2018,Peel2018,harnois_deraps2021}, 3pt correlation function \cite{heydenreich2022}. Here, we have extended the analysis in order to respond to the question about {\it which} HOS is most sensitive to MG deviations. To this end, we have used high order moments, the convergence PDF, peaks count in both the convergence and the aperture mass maps, and Betti numbers and Minkowski functionals as examples of topological indicators. In accordance with the literature and the common sense expectation, it turns out that all HOS probes are more efficient than the second order statistics, namely the convergence 2PCF, at spotting deviations from GR. One should however be careful in making the comparison fair by first cutting from the 2PCF data vector those scales which are cancelled out by the smoothing procedure applied to the convergence map before measuring the HOS data. Once this is done, we indeed find that the value of $f_{mod}({\cal{O}}, {\cal{M}})$, i.e., the number of configurations able to give a $3 \sigma$ detection of deviations from GR, is hardly different from the null value for 2PCF, while, on the contrary, it is possible to find at least one HOS probe with a positive value even for the weakest deviations caused by the fR6 model. However,  although there is some indication that $M_{\rm ap}$\,-\,peaks could be the favorite choice, it turns out that the answer to the question about {\it which HOS is best} has not a conclusive answer. Which HOS works best indeed depends on the source redshift, the model, and the specifics of the observational setup (e.g., the smoothing radius). This is not surprising given that each HOS probes the clustering properties of the underlying matter distribution in a different way being therefore more or less sensitive to deviations from GR depending on the scales where this takes place at any chosen redshift. For instance, considering the fR5 model, the 1\,-\,PDF is preferred over the MFs at all redshift but $z_s = 2$. However, adding neutrinos with $M_{\nu} = 0.10 \ {\rm eV}$ changes the relative ranking with MFs now performing better than 1\,-\,PDF also at $z_s = 1$. As a further example, we notice that the $M_{\rm{ap}}$\,-\,peaks typically provide the largest values of $f_{mod}({\cal{O}}, {\cal{M}})$, but this is no longer the case for the fR6$\_$0.10eV model whose signatures are traced in a similar way (if not better) by MFs. The lack of an absolute ranking of the HOS probes should however not be considered as a negative outcome, but rather as a good opportunity since it suggests that it is always possible to find a HOS probe able to discriminate among GR and MG theories. Such an intriguing possibility needs further test in order to be confirmed. Indeed, although a realistic noise has been added to the simulated maps, systematics are still missing. In particular, masking could affect in different ways topological and global estimators, while other effects such as photo\,-\,z uncertainties and shear multiplicative bias do propagate in different ways on each HOS probe. Moreover, when moving to the case with sources not at the same redshift, a sort of weighting of the results we find for single planes could take place, with the "weights" depending on the particular HOS. We plan therefore to revisit the {\it which} question in a forthcoming paper for the specific case of a Euclid\,-\,like survey.

Having decided {\it what} kind of theories we can test, and investigated {\it which} HOS we can use, it is still pending the question about {\it how} we decide whether an anomaly in the data is a statistically meaningful evidence for MG or a meaningless fluke due to noise in data from GR realizations. The naive answer would be to rely on the standard $\chi^2$, but there are actually a wide set of alternative tests which could be used. The motivation to go beyond the naive $\chi^2$ is twofold. On the one hand, we have introduced metrics which do not depend on the covariance so that they can also be used in cases where an accurate numerical estimate of this latter is computationally expensive. On the other hand, it is possible to make a metric more sensitive to small deviations by considering functions of the standard $\chi^2$ (e.g., the exponential entering the definition of Hellinger distance). We have therefore first considered metrics which can somewhat be seen as generalization of the $\chi^2$, or that propose different ways of quantifying the distance between GR and MG predictions. We have indeed found that, although all metrics perform well enough in particular cases, the Mahalanobis and the Hellinger distance stand out as preferred choices. This is reassuring given that these two quantities are the most similar to the usual $\chi^2$ statistics so that their implementation in inference pipelines is quite trivial. As an alternative approach, rather than using the full dataset, one could also search for particular features in it that are most sensitive to deviations from GR. This is the reason why we have relied on non parametric tests which compares the distribution of measurements in specific elements of the data vectors rather than their overall behavior. Such methods should be considered complementary to the metrics based ones, and used to drive the analysis of the data, indicating which features must be focused on being the more promising for GR vs MG discrimination. Again, much work is needed to translate these results into requirements on the noise, hence the survey setup (in terms of shape noise, area, source density) needed to extract the most information from HOS data.

Earlier works sharing the same philosophy of this paper, such as \cite{Peel2018, davies2024}, have shown the potential of HOS in discriminating and constraining MG models. Here we presented what can be considered an extension of such analyses, having included a larger number of probes, together with the effects of massive neutrinos and realistic shape noise, compatible with Stage IV surveys requirements. Our results are in accordance with previous ones, especially regarding the performance of $M_{\mathrm{ap}}$\,-\,peaks. However, we have here shown that, for each given probe, the performances may critically depend on the observational setup (i.e., the range, binning, smoothing). In the future, we deem it would be worth investigating even more statistics, such as voids \cite{contarini2021_voids, maggiore2025}, which, with their ability to probe under-dense regions, where $f(R)$ effects are less screened, could represent another tool to break cosmological degeneracies.

The present paper could be considered as a preparatory work for addressing the problem of how to use HOS to discriminate among GR and MG theories. We have answered to the {\it what, which, how} questions that are needed before dealing with the actual data. Armed with these answers and with Euclid data incoming, it is now the moment to move from {\it we could do} to {\it we have done}. What was the future of cosmic shear is ready to become the present, so time for the lensing community to get ready for it.

\acknowledgments

AV and VFC acknowledge funding by the Agenzia Spaziale Italiana (\textsc{asi}) under agreement no. 2024-10-HH.0 and by INFN/Euclid Sezione di Roma. FB acknowledges the support of Istituto Nazionale di Fisica Nucleare (INFN), iniziativa specifica QGSKY. SV acknowledges the funding of the French Agence Nationale de la Recherche for the PISCO project (grant ANR-22-CE31-0004). This article is based upon work from COST Action CA21136 Addressing observational tensions in cosmology with systematics and fundamental physics (CosmoVerse) supported by COST (European Cooperation in Science and Technology). We wish to thank the "Summer School for Astrostatistics in Crete" for providing training on the statistical methods adopted in this work. The work of AV was partially supported by the research grant number 2022E2J4RK “PANTHEON: Perspectives in Astroparticle and Neutrino THEory with Old and New messengers” under the program PRIN 2022 funded by the Italian Ministero dell’Università e della Ricerca (MUR).

\clearpage

\appendix

\section{Statistics ranking tabulated results}
\label{appendix:tabulated_results}

For convenience, we report in Tab.\,\ref{tab:maxcvm_bestprobe} and Tab.\,\ref{tab:fmod_bestprobes_multi}, respectively, the results of the CvM test and of $f_{mod}$ for different configurations, together with the probe (or probes) reaching the maximum value of the corresponding quantity. In Tab.\,\ref{tab:maxcvm_bestprobe}, we restrict our attention to the smoothing scale $\theta_s = 2.4'$, since, as discussed in the text, it provides the best overall performance. We remind the reader that the value of the test statistic is reported only for the cases in which $p < \alpha$.

\begin{table}[ht]
\centering
\caption{CvM results and best-performing probe by cosmology and source redshift, with $\theta_s = 2.4'$.}
\label{tab:maxcvm_bestprobe}
\begin{tabular}{l c S l}
\toprule
\textbf{Cosmology} & $\boldsymbol{z_s}$ & {\textbf{Max CvM}} & \textbf{Best Probe} \\
\midrule
fR4 & 0.5 & 49.004975 & $V_1$ \\
fR4 & 1.0 & 29.930191999999977 & $V_0$ \\
fR4 & 2.0 & 24.216534999999965 & $V_1$ \\
fR4 & 4.0 & 20.369527000000005 & $1\,-\,$PDF \\
fR5 & 0.5 & 20.87050099999999 & $\kappa$-peaks \\
fR5 & 1.0 & 24.45677599999999 & $V_2$ \\
fR5 & 2.0 & 46.70224999999999 & $\beta_1$ \\
fR5 & 4.0 & 9.01795599999997 & $1\,-\,$PDF \\
fR6 & 0.5 & 23.56614399999995 & $\beta_1$ \\
fR6 & 1.0 & 12.140457999999967 & $\beta_0$ \\
fR6 & 2.0 & 30.841876999999954 & $\langle \kappa^2 \rangle$ \\
fR6 & 4.0 & 50.21612099999999 & $\langle \kappa^2 \rangle$ \\
fR4\_0.3eV & 0.5 & 29.225393999999994 & $M_{\mathrm{ap}}$-peaks \\
fR4\_0.3eV & 1.0 & 18.311306000000002 & $V_2$ \\
fR4\_0.3eV & 2.0 & 15.55100699999997 & $M_{\mathrm{ap}}$-peaks \\
fR4\_0.3eV & 4.0 & 6.068097999999964 & $M_{\mathrm{ap}}$-peaks \\
fR5\_0.1eV & 0.5 & 27.84550299999995 & $\langle \kappa^4 \rangle$ \\
fR5\_0.1eV & 1.0 & 49.75247299999995 & $\beta_1$ \\
fR5\_0.1eV & 2.0 & 21.048438999999973 & $\langle \kappa^4 \rangle$ \\
fR5\_0.1eV & 4.0 & 36.18170199999997 & $\beta_1$ \\
fR5\_0.15eV & 0.5 & 65.43453399999999 & $\langle \kappa^4 \rangle$ \\
fR5\_0.15eV & 1.0 & 58.28153599999996 & $\langle \kappa^4 \rangle$ \\
fR5\_0.15eV & 2.0 & 52.40088399999996 & $\langle \kappa^4 \rangle$ \\
fR5\_0.15eV & 4.0 & 35.06829899999997 & $\langle \kappa^4 \rangle$ \\
fR6\_0.1eV & 0.5 & 35.445223999999996 & $M_{\mathrm{ap}}$-peaks \\
fR6\_0.1eV & 1.0 & 31.283491999999967 & $\langle \kappa^4 \rangle$ \\
fR6\_0.1eV & 2.0 & 29.561628999999982 & $1\,-\,$PDF \\
fR6\_0.1eV & 4.0 & 16.922071999999957 & $\langle \kappa^4 \rangle$ \\
fR6\_0.06eV & 0.5 & 44.94259925 & $M_{\mathrm{ap}}$-peaks \\
fR6\_0.06eV & 1.0 & 16.396102999999982 & $\langle \kappa^4 \rangle$ \\
fR6\_0.06eV & 2.0 & 14.775390999999956 & $\langle \kappa^4 \rangle$ \\
fR6\_0.06eV & 4.0 & 22.28533299999998 & $\kappa$-peaks \\
\bottomrule
\end{tabular}
\end{table}

\begin{table}[ht]
\centering
\caption{Max $f_{mod}$ and best-performing probe(s) by cosmology and source redshift.}
\label{tab:fmod_bestprobes_multi}
\begin{tabular}{l c S l}
\toprule
\textbf{Cosmology} & $\boldsymbol{z_s}$ & {\textbf{Max $f_{mod}$}} & \textbf{Best Probe(s)} \\
\midrule
fR4 & 0.5 & 100.0 & $1\,-\,$PDF, $\kappa\,-\,$peaks, $M_{\rm{ap}}\,-\,$peaks \\
fR4 & 1.0 & 100.0 & $1\,-\,$PDF, $\kappa\,-\,$peaks, $M_{\rm{ap}}\,-\,$peaks, BNs, MFs \\
fR4 & 2.0 & 100.0 & $1\,-\,$PDF, $M_{\rm{ap}}\,-\,$peaks, BNs, MFs \\
fR4 & 4.0 & 100.0 & $1\,-\,$PDF, $M_{\rm{ap}}\,-\,$peaks, BNs \\
fR4\_0.3eV & 0.5 & 20.0 & $1\,-\,$PDF, MFs \\
fR4\_0.3eV & 1.0 & 16.0 & $1\,-\,$PDF \\
fR4\_0.3eV & 2.0 & 54.6 & MFs \\
fR4\_0.3eV & 4.0 & 48.0 & $1\,-\,$PDF \\
fR5 & 0.5 & 100.0 & $1\,-\,$PDF, $M_{\rm{ap}}\,-\,$peaks \\
fR5 & 1.0 & 100.0 & $1\,-\,$PDF, $M_{\rm{ap}}\,-\,$peaks \\
fR5 & 2.0 & 100.0 & $M_{\rm{ap}}\,-\,$peaks \\
fR5 & 4.0 & 52.2 & MFs \\
fR5\_0.10eV & 0.5 & 100.0 & $1\,-\,$PDF \\
fR5\_0.10eV & 1.0 & 96.0 & $1\,-\,$PDF \\
fR5\_0.10eV & 2.0 & 80.9 & $M_{\rm{ap}}\,-\,$peaks \\
fR5\_0.10eV & 4.0 & 30.4 & MFs \\
fR5\_0.15eV & 0.5 & 44.0 & $1\,-\,$PDF \\
fR5\_0.15eV & 1.0 & 56.0 & $1\,-\,$PDF \\
fR5\_0.15eV & 2.0 & 100.0 & $M_{\rm{ap}}\,-\,$peaks \\
fR5\_0.15eV & 4.0 & 100.0 & $M_{\rm{ap}}\,-\,$peaks \\
fR6 & 0.5 & 72.0 & $1\,-\,$PDF \\
fR6 & 1.0 & 56.0 & $1\,-\,$PDF \\
fR6 & 2.0 & 0.0 & $1\,-\,$PDF, $\kappa\,-\,$peaks, $M_{\rm{ap}}\,-\,$peaks, BNs, MFs \\
fR6 & 4.0 & 28.0 & $1\,-\,$PDF \\
fR6\_0.06eV & 0.5 & 4.0 & $1\,-\,$PDF \\
fR6\_0.06eV & 1.0 & 4.4 & MFs \\
fR6\_0.06eV & 2.0 & 0.0 & $1\,-\,$PDF, $\kappa\,-\,$peaks, $M_{\rm{ap}}\,-\,$peaks, BNs, MFs \\
fR6\_0.06eV & 4.0 & 8.0 & $1\,-\,$PDF \\
fR6\_0.10eV & 0.5 & 4.0 & MFs \\
fR6\_0.10eV & 1.0 & 43.5 & MFs \\
fR6\_0.10eV & 2.0 & 52.4 & $M_{\rm{ap}}\,-\,$peaks \\
fR6\_0.10eV & 4.0 & 38.1 & $M_{\rm{ap}}\,-\,$peaks \\
\bottomrule
\end{tabular}
\end{table}


\bibliography{modgrav}
\bibliographystyle{JHEP} 



\end{document}